\documentclass[runningheads]{templates/llncs}
\usepackage{graphicx}
\usepackage[colorlinks=True,hypertexnames=false]{hyperref}
\usepackage{amsmath}
\usepackage{amssymb}
\usepackage{wasysym} 
\usepackage{soul} 
\usepackage{xcolor}
\usepackage{cancel} 
\usepackage{lscape}
\usepackage[left=0.7in,right=0.7in,top=1in,bottom=0.7in]{geometry} 
\usepackage{upgreek} 
\usepackage{booktabs} 
\usepackage{tabularx} 
\usepackage{outlines} 
\usepackage{enumitem}
\usepackage{amsmath}
\usepackage{amsfonts}
\usepackage{svg}
\usepackage{amssymb}

\usepackage{amsthm}
\usepackage{xcolor}
\usepackage{comment}
\usepackage{algorithm2e}
\usepackage{booktabs}
\usepackage{multirow}
\usepackage{lineno}

\makeatletter
\def\@fnsymbol#1{\ensuremath{\ifcase#1\or\star\or\dagger\or\ddagger\or\mathsection\or\mathparagraph\else\@ctrerr\fi}}
\makeatother

\usepackage{setspace}
\usepackage[backend=bibtex,giveninits=true,style=numeric,citestyle=numeric, doi=false,isbn=false,url=false,eprint=false,maxbibnames=4]{biblatex}
\bibliography{refs}
\setlist{topsep=1em, itemsep=0.5em}

\newtheoremstyle{customdef}
  {3pt} 
  {3pt} 
  {\itshape} 
  {} 
  {\bfseries} 
  {.} 
  { } 
  {} 

\theoremstyle{customdef}

\title{Contact, conflict, or opportunity? Out-group exposure creates tie opportunity, not tolerance}

\author{Mauritz N. Cartier van Dissel\thanks{These authors contributed equally.}\inst{1}\inst{2} \and%
Tom{\'a}{\v{s}} Lintner$^{\star\dagger}$\inst{3}\inst{4} \and Samuel Martin-Gutierrez \inst{5}\inst{1}\inst{2} \and%
Fariba Karimi$^{\dagger}$\inst{2}\inst{1}}

\authorrunning{M. N. Cartier van Dissel, T. Lintner, S. Martin-Gutierrez and F. Karimi}

\institute{Network Inequality Group, Complexity Science Hub, Vienna, Austria
\and
Institute of Human-Centred Computing, Graz University of Technology, Graz, Austria
\and
Department of Educational Sciences, Faculty of Arts, Masaryk University, Brno, Czech Republic
\and
National Institute SYRI, Brno, Czech Republic
\and
Grupo de Sistemas Complejos, ETS de Arquitectura de Madrid, Universidad Politécnica de Madrid, Madrid, Spain
}

\begin{document}

\maketitle
\begingroup
\renewcommand{\thefootnote}{\fnsymbol{footnote}}
\footnotetext[2]{Corresponding authors.}
\endgroup

\begin{abstract}
   Three theories offer competing predictions about how people respond to growing diversity in their social environment. Contact theory suggests more exposure to out-groups reduces prejudice; conflict theory predicts a stronger in-group preference; structural opportunity theory argues that shifts in behaviour only reflect changes in the opportunity structure rather than in underlying preference. We test these predictions using friendship and rejection nominations from nearly 5,000 students in 228 classrooms, across gender, ethnicity, and socio-economic status. We estimate individual preference using a multilevel model based on the Wallenius hypergeometric distribution, which accounts for the finite, asymmetric pool of potential ties. Results show that for ethnicity and socio-economic status, preferences are largely unaffected by classroom composition. For gender, however, same-gender preference strengthens as the out-group increases, supporting conflict theory. This means greater diversity does not necessarily change the intrinsic preference of students toward out-group peers, but creates more opportunities for cross-group interactions.
\end{abstract}

\keywords{peer social networks, contact theory, conflict theory, inter-group relations}
%

\section{Introduction}

Human societies are structured by patterns of social connection that shape access to resources and opportunities. From early peer interactions to adult professional networks, the formation of social ties influences the distribution of information, emotional support, social capital, and power. A central mechanism underlying these patterned connections is homophily---the tendency for individuals to form relationships with others who share similar social characteristics \cite{mcpherson2001birds}. Homophily contributes to the emergence of social clustering and segregation across multiple domains, including educational settings, workplaces, neighbourhoods, and civic life, thereby playing a fundamental role in the reproduction of social inequality and intergroup divisions \cite{mcpherson2001birds, blau1977inequality,karimi2018homophily}. Understanding when and why group-based social boundaries harden or soften is therefore essential for explaining broader patterns of social integration and stratification. Despite its ubiquity, the processes through which demographic composition shapes homophily remain theoretically contested. A longstanding debate concerns whether increasing diversity in social environments weakens social boundaries by fostering intergroup contact and integration or instead sharpens divisions by activating perceptions of threat, competition, and defensive boundary work \cite{allport1954nature, blumer1958race}. While contact-based perspectives emphasise the potential of sustained intergroup interaction to reduce prejudice and facilitate cross-group affiliation \cite{pettigrew1998intergroup, pettigrew2011recent}, conflict-oriented approaches highlight the ways in which changing demographic balances may intensify group positioning and reinforce social divisions \cite{blumer1958race}. Complementing these preference-based accounts, structural theories stress that observed patterns of homophily may also arise from the demographic organisation of social environments, as the distribution of potential interaction partners shapes opportunities for tie formation independently of underlying preferences \cite{blau1977inequality, mcpherson2001birds}. Together, these perspectives offer distinct predictions regarding how diversity influences the structure of social networks.

Contact theory suggests that exposure to members of different social groups can foster more positive intergroup relations. In its classical formulation, sustained intergroup contact---particularly under conditions of equal status, cooperation, shared goals, and institutional support---reduces prejudice and social distance, thereby facilitating cross-group affiliation \cite{allport1954nature}. Subsequent theoretical developments have elaborated the psychological mechanisms through which contact operates, including reduced intergroup anxiety, increased empathy, and the formation of affective ties such as friendship \cite{pettigrew1998intergroup, pettigrew2011recent, hewstone2014contact}. Empirical research across a variety of social settings has shown that opportunities for meaningful intergroup interaction can be associated with more inclusive social relations and greater acceptance of out-group members \cite{boda2023ethnic, hajdu2021ethnic, hjalmarsson2023not, kawabata2011significance, quillian2003beyond}. At the same time, recent work emphasises that diversity alone is insufficient: individuals embedded in diverse environments do not necessarily form diverse networks, and contextual conditions such as institutional norms and social climates shape whether structural diversity translates into actual intergroup ties \cite{cikara2022hate, chan2025promoting, witkow2025perceptions, mckeown2025understanding}.

Conflict theory offers a contrasting account rooted in sociological theories of group position and perceived threat. From this perspective, changing demographic balances may intensify rather than attenuate social boundaries. In classical formulations, prejudice reflects a collective sense of group position in which groups react defensively to perceived challenges to their symbolic status, social dominance, or access to valued resources \cite{blumer1958race}. As minority presence increases or group balances shift, intergroup competition may become more salient, leading to stronger in-group preference, avoidance of out-group members, and resistance to social integration. Empirical studies consistent with this perspective have documented contexts in which greater diversity is associated with heightened segregation or reduced intergroup contact \cite{bellmore2007influence, joyner2000school, lintner2023ukrainian, smith2016ethnic}. More recent scholarship likewise stresses that intergroup contact should not be treated as a universally beneficial mechanism, as its social consequences depend on broader relational and institutional contexts \cite{mckeown2025understanding}.

A third perspective, structural opportunity theory, shifts attention from changing preferences to the demographic organisation of social environments. From this viewpoint, observed patterns of homophily may arise mechanically from the distribution of potential interaction partners, even when underlying preferences remain stable. Building on macrosociological theories of social structure, demographic composition shapes the probability of intergroup association by altering the opportunity set within which relationships are formed \cite{blau1977inequality}. Subsequent research in network sociology has emphasised the importance of distinguishing between baseline homophily driven by opportunity structures and inbreeding homophily reflecting preference beyond opportunity \cite{mcpherson2001birds,sajjadi2024unveiling,currarini2009economic}. Recent empirical work demonstrates that compositional features such as the consolidation of social attributes within local social settings can substantially increase observed segregation without necessarily implying stronger same-group preferences \cite{boller2025friendship}. Likewise, contemporary multilevel analyses highlight the continuing importance of structural opportunities for cross-group interaction in shaping friendship formation \cite{meleady2026longitudinal}.

These perspectives thus offer distinct predictions regarding how demographic composition shapes social networks. Structural opportunity theory predicts that compositional variation will alter observed segregation even if underlying social preferences remain constant. From this viewpoint, higher proportions of same-group peers increase the likelihood of forming in-group ties simply because they expand the pool of available interaction partners \cite{blau1977inequality, mcpherson2001birds}. Contact theory predicts that greater exposure to out-group members should weaken same-group preference and facilitate cross-group ties by reducing social distance and fostering positive intergroup experiences \cite{allport1954nature, pettigrew1998intergroup}. Conflict theory, by contrast, predicts that demographic change may intensify perceived threat, boundary defense, and social distancing, leading to stronger in-group preference and avoidance of out-group members as group balances shift \cite{blumer1958race}.

\begin{figure}[h]
    \centering
    \includegraphics[width=\linewidth,trim=0cm 0cm 0cm 0.0cm,clip]{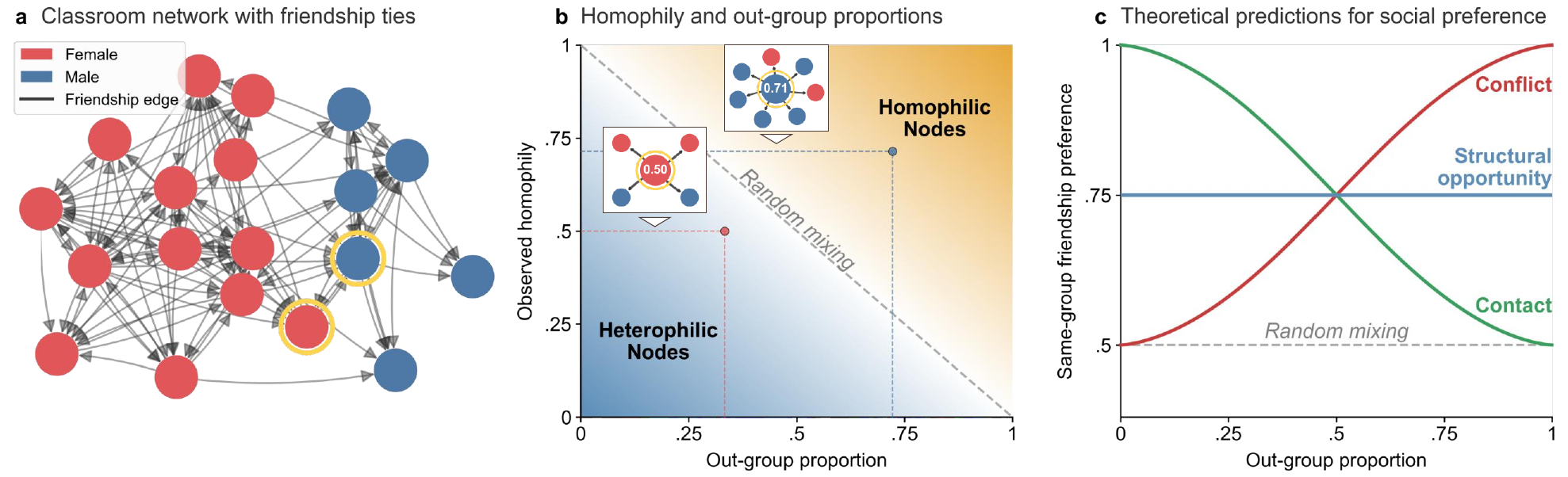}
    \caption{\textbf{Expected behaviour of observed homophily and preferences under the three theories for friendship ties.} Panel \textbf{a} shows a representative example of a friendship network from one of the classrooms in the analysed dataset, highlighting two focal nodes. Panel \textbf{b} plots observed homophily as a function of the proportion of out-group peers. The dashed line, which corresponds to random mixing, separates homophilic from heterophilic observations. The two highlighted nodes from panel (a) are marked, with their corresponding out-group proportions given by the share of out-group peers in their local environment (6/18 and 13/18, respectively). Panel \textbf{c} shows the stylised predictions of the three theories---structural opportunity, contact theory, and conflict theory---with respect to the underlying preference measure, illustrating how each theory implies a distinct relationship between preference and out-group proportion. For illustration, this panel focuses on the homophilic regime (preference $>0.5$). 
    }
    \label{fig:pipeline}
\end{figure}

A key implication of this theoretical debate is the need to distinguish between observed homophily and underlying intergroup preference. Rather than relying on network-level aggregate measures \cite{currarini2009economic,sajjadi2024unveiling}, which average out individual heterogeneity and may not be representative of any particular actor \cite{peel2018multiscale, k2025homophily,altenburger2018monophily}, we adopt individual-level measures of both homophily and preference to model behaviour at the level at which decisions are made.  We define observed homophily---also termed baseline homophily \cite{mcpherson2001birds}---as the proportion of same-group ties among all realised ties in an individual's local neighbourhood \cite{evtushenko2021paradox}.  Figure \ref{fig:pipeline}a shows a representative network from our dataset, where directed links represent declared friendships between students and nodes are coloured by gender. Two nodes are highlighted to illustrate the observed homophily metric in Figure \ref{fig:pipeline}b.

By construction, observed homophily conflates preference with opportunity: under random mixing, an out-group representing 70\% of the population is expected to occupy 70\% of an individual's neighbourhood. As a result, it mechanically declines as the out-group share increases (dashed line in Figure~\ref{fig:pipeline}b). An individual whose homophily exceeds this expectation is classified as homophilic; one whose homophily falls below it as heterophilic. What distinguishes the three theories, then, is not their expectations about observed homophily, which all agree will vary with group size, but their predictions about the preference that remains once demographic composition is accounted for. 

Figure~\ref{fig:pipeline}c illustrates the expected relationship between group composition and underlying preference under the three theoretical perspectives. Contact theory (green line) implies that as the relative presence of the out-group increases, same-group preference should weaken and cross-group preference should strengthen. Conflict theory (red line) predicts the opposite pattern: increasing out-group presence should heighten perceived competition or threat, resulting in stronger in-group preference and reduced cross-group affiliation. Structural opportunity theory (blue line) predicts that while observed homophily will mechanically increase with higher in-group proportions, underlying preference should remain stable once the opportunity structure is taken into account.

A central conceptual distinction in this study concerns how intergroup preference becomes observable in network behaviour. Rather than treating preference as a purely attitudinal or latent psychological disposition, we conceptualise in- and out-group preference as manifested through relational choices. In this behavioural sense, preference is reflected in whether individuals disproportionately direct interactions toward similar or dissimilar others once the opportunity structure of the local environment is taken into account. This conceptualisation aligns with network-analytic approaches that interpret tie selection as an observable expression of in- and out-group preference \cite{mcpherson2001birds,lee2019homophily}.

Intergroup preferences, however, may manifest quite differently depending on whether one examines positive or negative relational ties. Positive ties such as friendships often develop gradually through everyday interaction, shared activities, and the building of mutual trust. Negative ties---such as social rejection, avoidance, or dislike---may arise more quickly in response to perceived threat, competition, or the need to protect group boundaries \cite{hewstone2014contact}. This distinction matters because the same demographic conditions may foster positive connections between groups while having distinct, and not necessarily opposite, effects on negative ties.  Thus, homophily tends to be stronger in relationships that are more emotionally intense or socially consequential \cite{kretschmer2025strong}, and weaker when faced with rejection. 
One key contribution of this work is to analyse both positive ties and negative ties when assessing the effects of diversity and composition. 

Educational settings provide a particularly powerful empirical arena for examining these broader mechanisms. Classrooms constitute densely interconnected social environments in which demographic composition varies substantially across local contexts while interaction opportunities remain relatively structured. This makes them well suited for testing how group proportions shape both positive and negative peer relations. Moreover, dimensions such as gender, ethnicity, and socio-economic status remain highly salient in adolescent social organisation. Recent research indicates that the alignment of gender with other classroom characteristics can substantially amplify friendship segregation \cite{boller2025friendship}, that diversity does not automatically translate into inclusive peer relations \cite{chan2025promoting, witkow2025perceptions}, and that similarity in socio-economic background may strengthen peer influence processes relevant to educational inequality \cite{hovestadt2025similarity}. These findings suggest that classroom composition may influence peer networks through multiple pathways and that the relative importance of these pathways may vary across social dimensions and relational valence.

The present study leverages this empirical setting to investigate which of the three leading theories---contact, conflict, or structural opportunity---best accounts for the relationship between classroom composition and in- and out-group preference. Using harmonised data from seven independent studies conducted in Czech lower-secondary schools---including a nationally representative sample of sixth-grade classrooms---we investigate whether compositional variation in gender, ethnicity, and socio-economic status moderates patterns of in-group and out-group ties. To distinguish between opportunity-based and preference-based mechanisms, we compare observed homophily, capturing observed segregation, with a measure of local preference, which adjusts for the pool of available interaction partners. By integrating network-analytic approaches with cross-study inference, this study aims to provide a more generalisable test of whether demographic composition shapes social networks primarily through processes of contact, conflict, or structural opportunity, operating across friendship and rejection ties.

%
\section{Results}

\subsection*{Observed homophily patterns change across attributes}

\begin{figure}[h]
    \centering
    \includegraphics[width=0.85\linewidth,trim=0cm 0cm 0cm 0.18cm,clip]{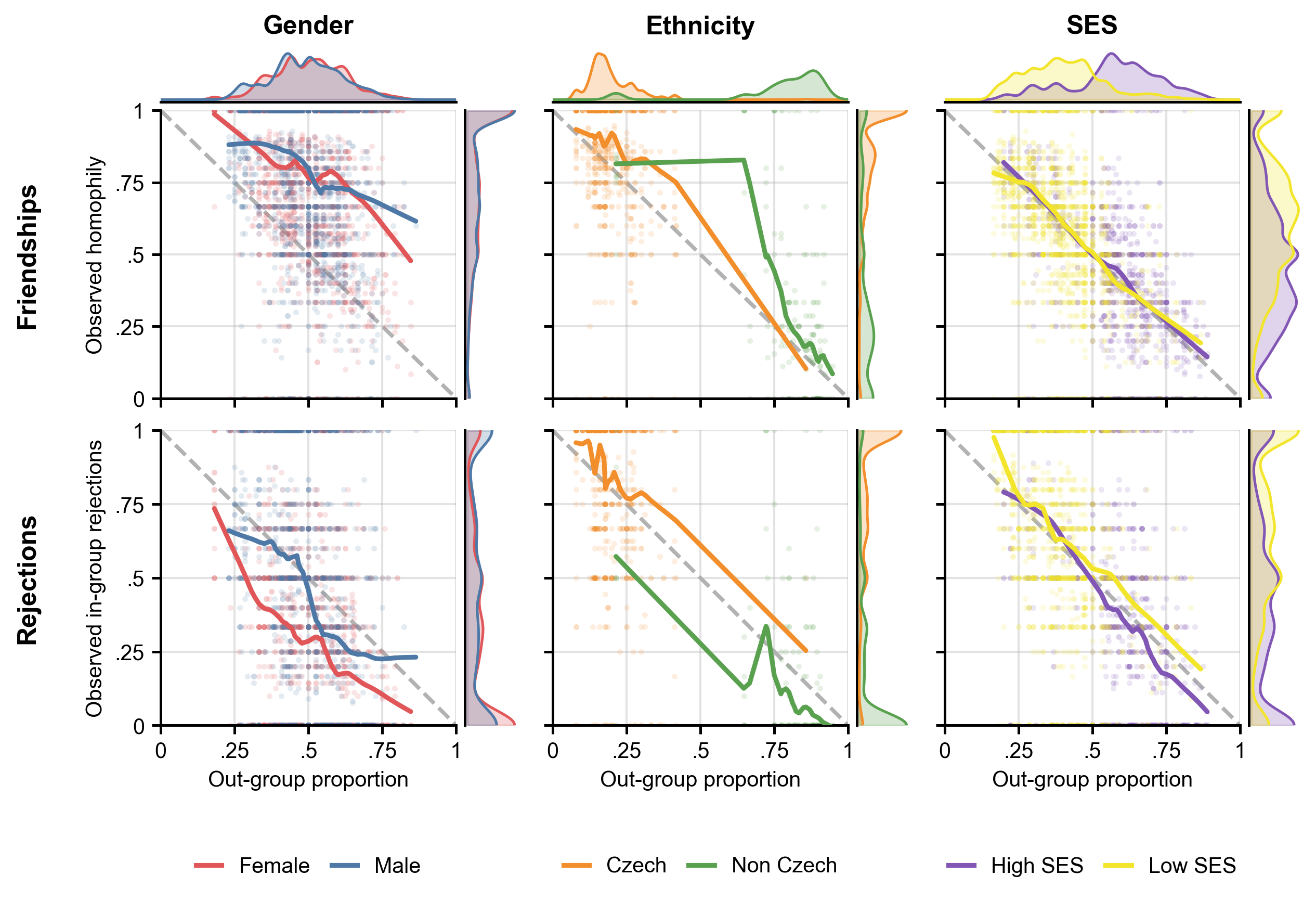}
    \caption{\textbf{Observed homophily as a function of out-group proportion.} Each panel shows individuals’ observed proportion of in-group ties against the proportion of out-group peers in the classroom. The top row shows friendship networks, and the bottom row shows rejection networks, while each column represents a different attribute. Points represent individual observations; solid lines indicate LOWESS smoothed trends by subgroup, and the dashed diagonal line represents the expected level of homophily under random mixing. Across all attributes, observed homophily decreases as out-group proportion increases, reflecting structural opportunity constraints. Gender exhibits strong homophily in friendships and structured patterns in rejection, while ethnicity shows weaker and noisier patterns, partly due to limited variation in classroom composition. SES closely follows the random-mixing baseline, suggesting lower salience in tie formation.}
    \label{fig:observed_homophily}
\end{figure}

Figure~\ref{fig:observed_homophily} shows the relation between observed homophily and out-group proportion for all three attributes---gender, ethnicity, and socio-economic status---and for both friendship and rejection ties. The scatterplot displays individuals’ observed proportion of in-group ties, while the solid line represents a locally weighted regression (LOWESS \cite{cleveland1981lowess}). For all attributes and for both friendship and rejection, we see a uniform pattern: with increasing out-group proportion, observed homophily decreases. This pattern is consistent with the expected baseline (dashed line), as a larger out-group proportion necessarily shrinks the pool of available in-group alters, mechanically reducing the opportunity to form in-group ties. Notice that observed homophily for rejection ties indicates the proportion of in-group rejections; therefore, a value below the baseline indicates more dislike for the out-group, while a value above the baseline indicates a greater tendency toward in-group dislike. Because the term `homophily' typically carries a positive connotation associated with friendship, we refer to this metric in the plots as `observed in-group rejections', although the underlying mathematical formula remains identical.

Clear differences arise across attributes. Gender exhibits strong homophily in friendship networks, as same-gender ties are more prevalent than expected under random mixing, evidenced by the majority of observations lying above the baseline. Notably, this pattern of homophily emerges because many students nominate exclusively same-group friends, leading to a substantial number of homophily values at 1, as indicated by the marginal density plots. Conversely, cross-gender rejections occur more frequently than expected, indicating that also in this case students perceive people from the same group more positively. For ethnicity, we observe modest homophily in friendships, although rejection ties display greater variability and less systematic structure. Note that we have limited data on ethnic classroom composition, as most Czech classrooms comprise a majority of Czech students, leaving us with limited data on classrooms where Czech students are not the majority. Finally, socio-economic status shows a pattern that is closely aligned with the expectation under random mixing. This suggests that SES is a comparatively less salient dimension in shaping both friendship and rejection ties within our study. 

These results demonstrate that as the out-group proportion increases, observed homophily diminishes across all social dimensions and for both types of ties. Yet this consistent trend points in theoretically opposite directions depending on the tie valence. For friendship ties, declining same-group nominations as the out-group grows would superficially suggest contact theory: students in more diverse classrooms appear to be forming more cross-group friendships, consistent with the idea that greater intergroup exposure fosters positive relations. For rejection ties, however, the same mechanical pattern tells a different story---a declining proportion of in-group rejections implies that an increasing share of hostility is directed outward toward the out-group, which would superficially align with conflict theory's prediction that a growing out-group triggers defensive boundary work and outward hostility. Crucially, both readings rest on the same arithmetic: as the out-group expands, it necessarily absorbs a larger share of whatever nominations a student makes, whether friendly or hostile. Because the reduction in same-group ties may therefore simply be a mechanical result of a shrinking pool of in-group peers, these observations alone cannot tell us whether underlying preferences have actually shifted in either direction. For this reason, to understand which theory truly has the most evidence in adolescent classrooms, we require a preference measure that removes the influence of the opportunity structure from our observations.

\subsection*{Modelling in-group preference and classroom composition}

To isolate underlying social preferences from the structural constraints of classroom size, we model the nomination process as a sequence of biased selections from a finite pool. Because classroom networks are small and students often nominate a substantial fraction of their peers, each choice alters the composition of the remaining pool. To account for this, we adopt a biased urn selection framework based on the Wallenius non-central hypergeometric distribution \cite{wallenius1964biased, fog2008sampling}. In this framework, in-group and out-group peers are assigned differential ``weights'', allowing us to estimate a normalised preference parameter ($q$) that remains invariant to the classroom's demographic proportions. On this scale, $q=0.5$ represents neutral mixing (treating all peers equally), $q>0.5$ indicates a genuine preference for one’s own group, and $q<0.5$ indicates a preference for the out-group. Intuitively, $q$ can be interpreted as the probability that a student would select an in-group peer if presented with a balanced one-versus-one choice. By removing the mechanical influence of the opportunity structure, this parameter allows us to discriminate between the competing predictions of contact, conflict, and structural opportunity theories at the level of individual preference. Further details of the model are provided in the Methods section.

To test whether in-group preference varies with classroom composition, we estimate a multilevel model that links individual tie formation to out-group proportion. The outcome of interest is the number of same-group nominations made by each student. Formally, for student $i$, the number of same-group nominations $s_i$ follows:

$$s_i \mid q_i \;\sim\; \text{WalleniusNCH}\!\left(S_i,\, N_i,\, k_i,\; q_i\right)$$
where $S_i$ is the number of available in-group peers, $N_i$ the total number of peers, and $k_i$ the number of nominations made. We transform the preference parameter  $q_i$ into a log-odds scale, $\eta_i=\text{logit}(q_i)$, to analyse how preferences shift across different classroom settings. Much like a standard binomial or logistic regression \cite{weisberg2005applied}, this transformation allows us to model social preference as a linear function while ensuring that our final estimates for $q$ remain constrained between 0 (total out-group preference) and 1 (total in-group preference). Specifically, we model:

$$\eta_i = \underbrace{\alpha_{g_i}}_{\text{intercept}} + \underbrace{\beta_{g_i} \cdot (p_{c_i} - 0.5)}_{\text{effect of out-group proportion}} + \underbrace{u_{c_i}}_{\text{classroom random effect}} + \underbrace{v_i}_{\text{student random effect}}$$ 
where $g_i$ and $c_i$ indicating the corresponding demographic group and classroom respectively, and $p_{c_i}=\frac{N_i-S_i}{N_i}$ the proportions of members of the out-group in classroom $c_i$. In this model, the $\alpha$ coefficient captures the overall homophily tendency, while the $\beta$ coefficient represents the slope of the relation between out-group proportion and preference. A positive $\beta$ indicates increasing homophily as the out-group size increases, a negative value suggests a decreasing relationship, whereas a $\beta$ of $0$ implies that there is no relation between preference and $p_{c_i}$. The random effects $u$ and $v$ account for heterogeneity at the classroom and individual levels, respectively. Both terms follow normal distributions centred at zero, with standard deviations $\sigma_\text{class}$ and $\sigma_\text{student}$. Including these effects is important, as research has shown that the empirical variability of attribute preferences frequently exceeds the expected variance \cite{altenburger2018monophily}.

It is worth noting that for positive and negative ties, both contact and conflict theory make opposite predictions regarding preferences. In friendship networks, conflict theory posits that a larger out-group intensifies in-group preference through perceived threat, whereas contact theory suggests that increased exposure reduces such bias. For rejection networks, these predictions flip: conflict theory expects hostility to be directed outward, manifesting as weaker in-group rejection (i.e., more out-group rejection) as the out-group grows. Conversely, contact theory anticipates that diversity fosters more inclusive social relations, thereby reducing out-group hostility and maintaining a relatively higher proportion of in-group rejections.

\subsection*{Evidence favours structural opportunity and conflict theories}

\begin{figure}[t]
    \centering
    \includegraphics[width=0.85\linewidth,trim=0cm 0cm 0cm 0.18cm,clip]{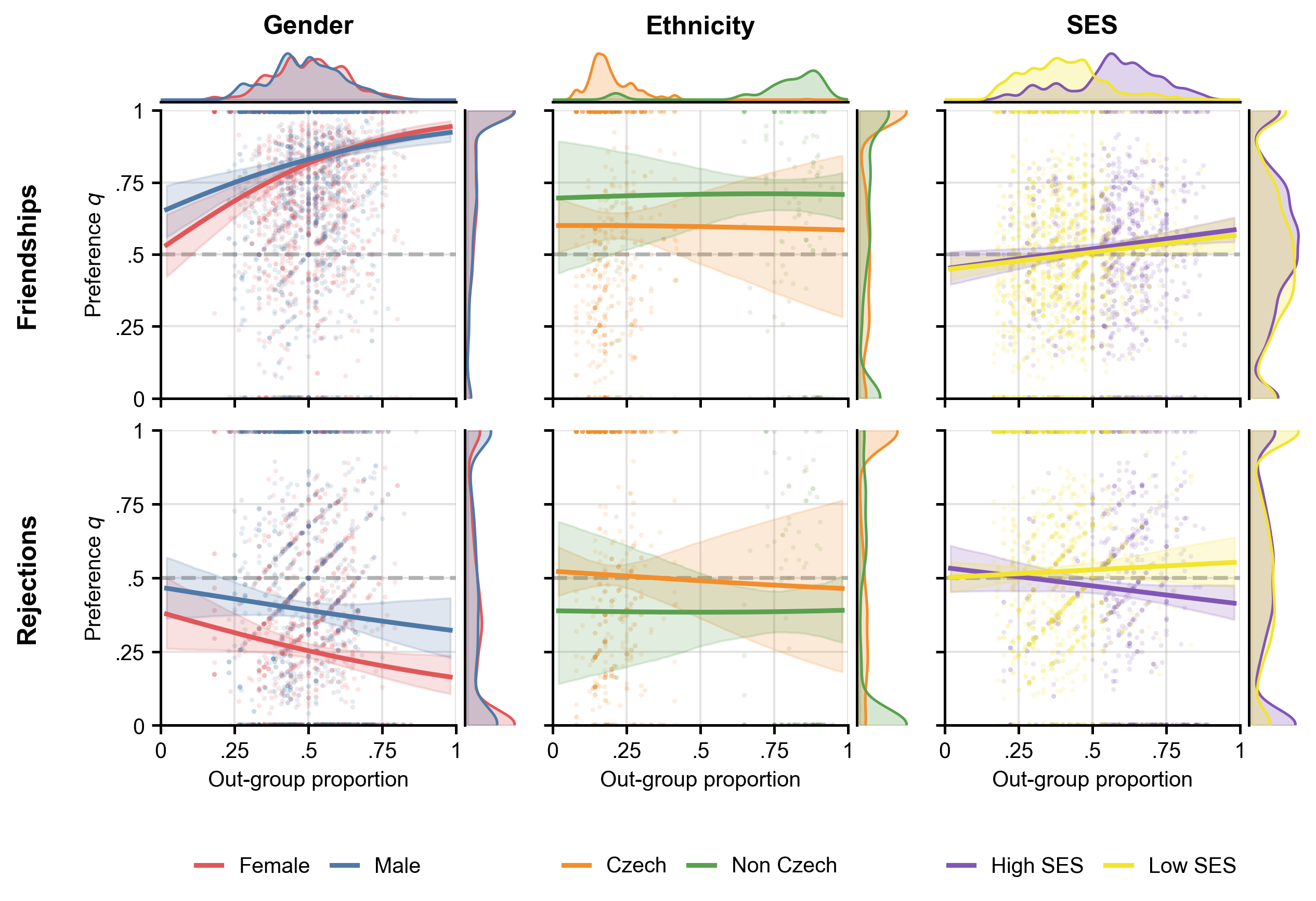}
    \caption{\textbf{Estimated in-group preference as a function of out-group proportion.} Each panel plots the estimated preference parameter $q$ against the proportion of out-group peers in the classroom, separately by attribute and link type. Points represent individual-level maximum likelihood estimates (MLEs) of $q$ obtained from the Wallenius model, while solid lines show fitted group-level relationships with shaded 90\% confidence intervals derived from the estimated uncertainty in $\beta$. The dashed horizontal line at $q=0.5$ corresponds to neutral mixing (no preference). Slopes capture deviations from structural opportunity: a flat relationship ($\beta=0$) indicates that observed behaviour is fully explained by compositional constraints, while positive or negative slopes indicate systematic preference shifts consistent with contact or conflict theory, depending on the link type. Across attributes, gender exhibits strong composition-dependent preference consistent with conflict theory, whereas ethnicity and SES largely align with the structural opportunity baseline.}
    \label{fig:preference}
\end{figure}

Table~\ref{tab:wallenius} reports the estimated baseline preferences ($q(\alpha)=\text{logit}^{-1}(\alpha)$), the effect of classroom composition ($\beta$), and the corresponding Posterior Model Probabilities (PMP; more details in the Methods section and the SI) for three competing hypotheses: structural opportunity theory ($\beta=0$), contact theory, and conflict theory. Figure~\ref{fig:preference} visualises the fitted relationships between out-group proportion and the estimated preference parameter $q$, with separate lines by group.

For gender, we find strong and statistically significant levels of homophily for friendship and out-group aversion for rejection ties across both males and females, as evidenced by the intercepts $q(\alpha)$. These results indicate a high intrinsic preference for same-group peers alongside a systematic dislike for the out-group. When examining how these preferences shift with classroom composition, the PMPs for friendship ties overwhelmingly favour the conflict-theory direction, with values near 1 for both female and male students. This strong theoretical support is reflected in positive and statistically significant estimated slopes (upper left panel in Figure \ref{fig:preference}), which indicate that as the out-group grows larger, students appear to intensify their preference for same-gender peers, over and above what structural opportunities alone would predict. For rejection ties, the evidence is weaker but broadly consistent with conflict theory’s reversed prediction. For females, the negative slope suggests that increasing out-group presence is associated with weaker in-group rejection preference (i.e., more rejection directed toward out-group peers), with moderate support for the conflict-theory direction. For males, however, the estimates are not very significant and the PMP favours the structural opportunity theory, indicating little systematic relationship between classroom composition and rejection behaviour.

For ethnicity, we find that statistically significant baseline homophily is present only for non-Czech students in friendship networks. While Czech students show a tendency toward same-group friendship and both groups tend to reject the out-group, the intercepts do not reach statistical significance. When examining how these preferences shift across different classroom settings, the results consistently favour structural opportunity theory. The PMPs assign the highest probability to the null model $(\beta=0)$ for all ethnic groups and tie types, with probabilities ranging from $0.57$ to $0.73$. These results are reflected in estimated slopes that are small and highly uncertain, with confidence intervals spanning zero in all cases. This indicates that observed patterns of same- and cross-ethnic ties are largely accounted for by the availability of peers, with little evidence of composition-dependent preference. Finally, it is important to note that many classrooms do not include non-Czech students, meaning we have considerably less data for this attribute; this likely contributes to the low statistical significance.

\begin{table}[t]
\centering
\small
\setlength{\tabcolsep}{6pt}
\begin{tabular}{lll r rr rrr rr}
\hline
Attribute & Link type & Group
  & q($\alpha$)
  & $\beta$ & SE$(\beta)$
  & PMP$_\text{contact}$ & PMP$_0$ & PMP$_\text{conflict}$
  & $\sigma_\text{class}$ & $\sigma_\text{student}$ \\
\hline

\multirow{4}{*}{Gender}
 & \multirow{2}{*}{Friendship}
 & Female    & $0.82^{***}$ & $2.84^{***}$ & $0.53$ & $<0.01$ & $<0.01$          & $\mathbf{>0.99}$ & \multirow{2}{*}{0.75} & \multirow{2}{*}{0.87} \\
 &           & Male        & $0.83^{***}$ & $1.94^{***}$ & $0.50$ & $<0.01$ & 0.01          & $\mathbf{0.99}$  &                       &                       \\
 & \multirow{2}{*}{Rejection}
 & Female    & $0.25^{***}$ & $-1.18$        & $0.62$ & 0.01    & 0.30             & $\mathbf{0.70}$  & \multirow{2}{*}{0.83} & \multirow{2}{*}{0.80} \\
 &           & Male        & $0.39^{***}$ & $-0.64$        & $0.54$ & 0.03 & $\mathbf{0.68}$  & 0.29 &                       &                       \\
\hline
\multirow{4}{*}{Ethnicity}
 & \multirow{2}{*}{Friendship}
 & Czech     & $0.60$       & $-0.04$        & $1.06$ & 0.10 & $\mathbf{0.57}$  & 0.34 & \multirow{2}{*}{0.63} & \multirow{2}{*}{0.43} \\
 &           & Non-Czech   & $0.71^{**}$  & $-0.02$        & $0.93$ & 0.17 & $\mathbf{0.72}$  & 0.12 &                       &                       \\
 & \multirow{2}{*}{Rejection}
 & Czech     & $0.49$       & $-0.28$        & $1.04$ & 0.10 & $\mathbf{0.71}$  & 0.18 & \multirow{2}{*}{0.50} & \multirow{2}{*}{0.42} \\
 &           & Non-Czech   & $0.38$       & $0.07$       & $1.03$ & 0.16 & $\mathbf{0.73}$  & 0.11 &                       &                       \\
\hline
\multirow{4}{*}{SES}
 & \multirow{2}{*}{Friendship}
 & High SES  & $0.52^{*}$  & $0.56^{*}$  & $0.25$ & $<0.01$    & 0.39          & $\mathbf{0.60}$  & \multirow{2}{*}{0.25} & \multirow{2}{*}{0.21} \\
 &           & Low SES     & $0.51$       & $0.49^{*}$  & $0.23$ & $<0.01$ & $\mathbf{0.50}$ & 0.49 &                       &                       \\
 & \multirow{2}{*}{Rejection}
 & High SES  & $0.47$       & $-0.49$        & $0.34$ & 0.01 & $\mathbf{0.67}$  & 0.32 & \multirow{2}{*}{0.41} & \multirow{2}{*}{$<0.001$} \\
 &           & Low SES     & $0.53^{*}$  & $0.21$       & $0.33$ & 0.10 & $\mathbf{0.86}$  & 0.04 &                       &                           \\

\hline
\
\end{tabular}
\caption{\textbf{Multilevel Wallenius Model Results.} This table reports the estimated baseline preferences, composition effects, and evidence for competing social theories across gender, ethnicity, and SES for both friendship and rejection ties. The intercept $q(\alpha)$ represents the baseline preference on the unit interval: $q>0.5$ indicates in-group preference, $q<0.5$ indicates out-group preference, and $q=0.5$ represents neutral mixing. The $\beta$ coefficient captures the slope of the relationship between out-group proportion and preference; a positive slope indicates increasing homophily as the out-group becomes more prevalent, while a negative slope suggests the opposite. Posterior Model Probabilities (PMP) are provided for three competing hypotheses: contact theory (PMP$_\text{contact}$), structural opportunity (PMP$_0$, representing the null where $\beta=0$), and conflict theory (PMP$_\text{conflict}$). Bold values indicate the theory with the highest posterior probability for each row. Random-effect standard deviations ($\sigma_\text{class}$ and $\sigma_\text{student}$) account for heterogeneity at the classroom and student levels. Significance stars indicate results from two-sided Wald tests on the Maximum Likelihood Estimates ($^{*}p<0.05,^{**}p<0.01,^{***}p<0.001$)}
\label{tab:wallenius}
\end{table}

For socio-economic status, we find that baseline preference levels for both friendship and rejection ties hover near $0.5$ (neutrality). Whilst there is marginal statistical significance for High SES friendship homophily and Low SES in-group rejection, the departure from the null is negligible. Regarding the impact of classroom composition, the Posterior Model Probabilities for friendship ties present a divided picture; whilst there is modest support for the conflict-theory direction for High SES students, the evidence is evenly split for Low SES students. These values are reflected in small negative slopes, suggesting that any preference-based response to SES composition is minimal. In rejection networks, the PMP results favour structural opportunity theory, with estimated slopes centred near zero for both groups. 

To account for variation beyond these group-level trends, the model includes random-effect standard deviations, $\sigma_\text{class}$ and $\sigma_\text{student}$, which capture unexplained heterogeneity at the classroom and individual levels, respectively. For gender, we find the highest levels of variability, indicating that both individual dispositions and local classroom climates significantly influence the baseline preference $q$. This suggests that gender dynamics are highly sensitive to specific peer-governed norms and unique contextual factors. In contrast, the much smaller sigma values for socio-economic status reveal far more uniform preference patterns across the sample. It is important to clarify that while a standard deviation of approximately $0.8$ may seem substantial if interpreted directly as a shift in the preference parameter $q$, its substantive impact is relatively moderate when one considers that these random effects are estimated in logit space \footnote{If $q(\alpha)=0.8$, then $\alpha=\operatorname{logit}(0.8)\approx 1.4$, so $\alpha\pm0.8=[0.6,2.2]$ maps to approximately $[0.64,0.90]$ on the probability scale.}.

Overall, our results show virtually no evidence to support contact theory within the social preferences of these adolescent classrooms. Across all social dimensions and both positive and negative ties, the maximum support found for the contact-theory direction was a negligible Posterior Model Probability of $0.17$, with most values falling far lower. This finding stands in stark contrast to the intuitive patterns observed in the raw data for friendships of Figure \ref{fig:observed_homophily}, where the consistent decline in same-group ties as the out-group grows might superficially suggest a natural process of intergroup integration.

%

\section{Discussion}

Our findings paint an asymmetric picture of how demographic composition shapes peer networks in lower-secondary classrooms. Across gender, ethnicity, and SES, observed homophily declines with out-group proportion at roughly the rate predicted by the compositional baseline \cite{blau1977inequality, mcpherson2001birds}. Yet once the opportunity structure is adjusted for, the three dimensions diverge sharply. For ethnicity and SES, preference is largely invariant to composition, consistent with structural opportunity theory. For gender, by contrast, same-gender preference in friendship \emph{intensifies} as the other-gender share in the classroom grows---a pattern consistent with conflict theory \cite{blumer1958race}.

Why should gender be the axis along which composition most sharply shapes preference? Three interlocking features of early-adolescent social organisation offer an explanation. First, gender is arguably the most perceptually salient and continuously enforced social category in this developmental window: unlike ethnicity or SES, which may be ambiguous, private, or contextually variable, gender is immediately visible, publicly marked through appearance and language, and continually reinforced by peers, teachers, and institutional routines \cite{maccoby1998two, martin2011patterns}. Second, early adolescence is a period of heightened gender-identity consolidation, during which peer affirmation of gender-typical behaviour is especially rewarding and deviation especially costly \cite{hill1983menstrual, galambos2009gender, mehta2009samegender}. Third, cross-gender friendships at this age carry distinctive social risks: they are easily reinterpreted as romantic, subject to teasing, and often sanctioned by same-gender peers \cite{mehta2009samegender}. Together, these features make gender a dimension along which out-group size is most plausibly experienced as threat rather than opportunity---precisely the configuration under which conflict theory predicts that in-group preference should intensify \cite{blumer1958race, hewstone2014contact}.

The classroom context helps clarify why Allport's contact conditions may be systematically harder to satisfy for gender than for other dimensions. Equal status, cooperative interdependence, shared goals, and institutional support \cite{allport1954nature, pettigrew1998intergroup} are more readily engineered around academic tasks---which cut across ethnic and class lines by design---than around informal peer interaction, where gender norms are peer-governed rather than teacher-governed \cite{maccoby1998two, thorne1993gender}. Classrooms therefore do not automatically function as neutral sites of contact for gender; compositional imbalance on an identity-laden, behaviourally enforced dimension appears instead to convert the classroom into an arena in which intergroup salience rises precisely when contact would, in principle, be most available. This reading aligns with a growing literature emphasising that structural diversity does not translate mechanically into relational integration without supportive institutional and normative conditions \cite{chan2025promoting, witkow2025perceptions, mckeown2025understanding}.

This conflict-theoretic pattern carries a concrete implication for minority-gender students. For a boy in a predominantly female classroom, or a girl in a predominantly male one, the combination of a small same-gender pool and an intensified same-gender preference produces a structural squeeze: in-group partners are mechanically limited, while the willingness to compensate through cross-gender ties is reduced, not enhanced, by the imbalance. The predicted consequence could be that minority-gender students form fewer friendships overall than their majority counterparts---not because the classroom offers fewer potential partners, but because the partners it offers are, in a motivational sense, devalued (more on this in Section \ref{SI_sec:outgroup_outdegree} of the SI). This mechanism may help reconcile contact-theoretic optimism about diversity with the repeated observation that greater compositional heterogeneity does not reliably improve cross-group relations \cite{bellmore2007influence, joyner2000school, smith2016ethnic}. 

The contrast with ethnicity and SES is instructive. For both dimensions and both tie valences, we find little evidence of preference-based deviation from the compositional baseline. Several non-mutually-exclusive explanations apply: ethnicity and SES are less continuously visible in classroom interaction than gender; Czech lower-secondary classrooms are predominantly ethnically homogeneous, limiting statistical power where conflict-theoretic predictions would be sharpest; and SES similarity may operate through subtler status processes not well captured by a binary partition \cite{hovestadt2025similarity}. The substantive conclusion is that, once opportunity is accounted for, students do not systematically steer toward or away from ethnic or class out-groups as composition varies---converging with recent evidence that consolidation and local environmental features, rather than preference shifts, drive much of what is conventionally labelled ethnic or class segregation \cite{boller2025friendship, meleady2026longitudinal}.

Friendship and rejection diverge in theoretically informative ways. For gender, the robust conflict-consistent signal in friendship is not mirrored in rejection: the female coefficient weakly aligns with the reversed conflict prediction, while the male coefficient is essentially flat. This is consistent with the view that affiliative and aversive ties follow different psychological logics---friendship requires cumulative investment and is sensitive to aggregate compositional pressure, whereas rejection can arise abruptly in response to isolated incidents and may be less sensitive to classroom-level balance \cite{hewstone2014contact, kretschmer2025strong}. 

Our approach introduces a novel methodological framework for testing intergroup theories by shifting the focus from aggregate network-level metrics to individual-level social preferences. This distinction addresses a long-recognised concern that aggregate summaries can obscure the heterogeneity that defines actor-level behaviour \cite{peel2018multiscale,altenburger2018monophily}, and operates at the level at which the three theories make competing predictions. The Wallenius non-central hypergeometric likelihood, embedded in a multilevel model with classroom and student random effects, explicitly accounts for the depletion of the available pool as nominations are made---a correction that matters since the same-group pool can be extremely small for minorities and students often nominate a substantial share of their peers. As shown in Section \ref{SI_sec:model-comparison} of the SI, the absence of support for contact theory is robust across specifications, including the binomial approximation \cite{altenburger2018monophily} and Fisher's non-central hypergeometric model \cite{fog2008sampling}, while the conflict-consistent signal for gender---particularly for boys---emerges most clearly when the finite pool is modelled explicitly.

Equally important is the inferential frame in which the three theories are compared. Because structural opportunity corresponds to a point null ($\beta = 0$), we use a spike-and-slab prior \cite{mitchell1988bayesian} that assigns explicit prior mass to the null and partitions the slab into the contact and conflict consistent directions, yielding three directly interpretable posterior model probabilities. This allows structural opportunity to be evaluated as a substantive theoretical position rather than as a hypothesis one merely fails to reject.

This work has a number of limitations that point toward important future directions. First, our measures of peer relations rely on bounded nomination procedures, which may not capture the full range of students' social ties, and missing data---particularly for ethnicity---limit statistical power in some analyses. Furthermore, our analysis of ethnicity is constrained by the demographic composition of Czech classrooms, which are predominantly ethnically homogeneous and typically feature a large Czech majority alongside a small non-Czech minority. Future research should focus on analysing more diverse ethnic scenarios (particularly those where groups occupy intermediate positions) in order to provide a more comprehensive picture of how ethnicity interacts with classroom composition. 

Second, although we examine three dimensions of social identity simultaneously, which already represents an advance over most single-attribute studies, we do not model their intersections. Students occupy multiple group memberships at once, and the compounding of, say, gender and ethnic minority status may produce preference dynamics that neither dimension alone would predict \cite{martin2025intersectional}. However, the three focal theories---contact, conflict, and structural opportunity---provide predictions only for binary in-group/out-group partitions, leaving it theoretically unclear how the mechanisms would behave in a multi-group setting. 
Future work translating the theories to an intersectional framework, similar to \cite{martin2026simple}, may substantially deepen our understanding of how composition shapes peer networks.

On the applied side, our findings raise important questions for school-level intervention. Compositional diversity alone appears insufficient to shift underlying preferences, particularly along gender lines---suggesting that efforts to enhance intergroup integration may need to move beyond simply mixing students and instead attend to the normative and institutional conditions under which cross-group interaction occurs. Future research should examine which sociodemographic dimensions are most responsive to structural intervention, and whether gender-sensitive approaches---such as those targeting the social costs of cross-gender friendship in early adolescence---can attenuate the conflict-consistent preferences we document here.

%
\section{Methods}

\subsection*{Data}

The study draws on seven independent datasets collected in lower-secondary school classrooms in the Czech Republic. An overview of the datasets is provided in Table~\ref{tab:datasets}, while detailed descriptions of sampling procedures, survey instruments, and data collection contexts are available in Section \ref{SI_sec:data_description} of the Supplementary Information. 

The datasets were selected for their comparability in terms of target population, measurement of peer relations, and availability of key sociodemographic indicators. Rather than aiming to estimate aggregate levels of homophily, the purpose of combining these datasets was to maximise variability in classroom demographic composition. This design allows us to approximate a natural experiment in which students are quasi-randomly allocated to classrooms with differing gender balances, ethnic compositions, and socio-economic distributions, thereby creating variation in local opportunity structures for tie formation. Because student assignment to classrooms is largely determined by administrative processes rather than individual choice, variation in classroom composition can be treated as plausibly exogenous to individual tie preferences.

All datasets are analysed cross-sectionally. Although several original studies employed longitudinal designs, only the first available wave is used here to ensure comparability of measurement timing and network structure.

The \textit{PanelSYRI} dataset \cite{tabery2024syri} provides a nationally stratified sample of sixth-grade classrooms across the Czech Republic. The \textit{Defending Victims of Bullying} dataset \cite{lintner2025affective} includes seventh-grade classrooms from Prague schools. The \textit{Classroom Discourse} dataset \cite{lintner2021classroom} covers ninth-grade classrooms in the South Moravian Region, while the \textit{Collectivity} dataset \cite{lintner2023collectivity} includes sixth-grade classrooms from the same region. The \textit{Ukrainian Refugees} dataset \cite{lintner2023ukrainian} comprises classrooms from grades five to nine in Brno schools. The remaining two datasets---\textit{Vietnamese in Czech Schools} \cite{konradova2012vietnamci} and \textit{Peer Group \& Participation} \cite{vaskova2019participace}---originate from publicly available undergraduate thesis projects conducted in Czech lower-secondary settings.

\begin{table}[ht]
\small
\centering
\caption{Overview of datasets included in the study}
\label{tab:datasets}
\begin{tabular}{l l c c c c c c c c l}
\hline
Dataset & Location & Grades & Classrooms & Students & Friendship & Rejection & Gender & Ethnicity & SES & Source \\
\hline
PanelSYRI & National & 6 & 134 & 2609 & X & X & X & X & X & \cite{tabery2024syri} \\
Defending Victims of Bullying & Prague & 7 & 39 & 910 & X & X & X & X & X & \cite{lintner2025affective} \\
Classroom Discourse & South Moravian Region & 9 & 21 & 435 & X & X & X & X & X & \cite{lintner2021classroom} \\
Collectivity & South Moravian Region & 6 & 12 & 276 & X &  & X & X & X & \cite{lintner2023collectivity} \\
Ukrainian Refugees & Brno & 5--9 & 12 & 266 & X & X & X & X & X & \cite{lintner2023ukrainian} \\
Vietnamese in Czech Schools & {\v C}esk{\'a} L{\'i}pa & 5--9 & 8 & 167 & X & X & X & X &  & \cite{konradova2012vietnamci} \\
Peer Group \& Participation & Svitavy \& Brno-Country & 5 & 2 & 34 & X & X & X &  &  & \cite{vaskova2019participace} \\
\hline
\end{tabular}
\end{table}

To ensure the robustness of our estimates, we applied several preprocessing criteria to the combined dataset. First, for each attribute-link combination, we only included classrooms containing at least three students per demographic group (e.g., at least three males and three females) to ensure that group-level dynamics were distinguishable. Furthermore, classrooms with more than 30\% missing data were excluded to prevent results from being skewed by the influence of non-responses. Finally, at the individual level, we only considered students who made at least one nomination whilst not having nominated every available peer, as such cases offer no information about the underlying social preference. A table indicating the number of students and classrooms considered for each tie-attribute combination can be found in Section \ref{SI_sec:number_samples} of the SI.

\subsection*{Observed homophily and structural constraints}
 
We start by defining the descriptive quantity shown in the homophily plots. For student $i$, let $k_i$ denote the total number of nominations made for a given tie type and let $s_i$ denote the number of those nominations that go to same-group classmates. We define \emph{observed homophily} as
\[
h_i = \frac{s_i}{k_i},
\]
that is, the fraction of realised ties that connect to alters from the student's own group. Notice that for negative ties, we refer to this metric as \emph{observed in-group rejections}. As shown in the initial visualisations in Figure \ref{fig:observed_homophily}, this measure is inherently confounded because it conflates underlying social preference with the structural opportunity afforded by the classroom's demographic composition. 

Let $S_i$ denote the number of available same-group classmates for student $i$ and let $N_i$ denote the total number of available classmates. Notice that because self-ties are not possible, the focal student is excluded, so $S_i$ and $N_i$ count only available alters. We therefore define
\[
p_i = \frac{N_i-S_i}{N_i},
\]
that is, the proportion of out-group classmates in student $i$'s opportunity set. Notably, the corresponding group-specific out-group proportions do not sum to one across complementary groups. To illustrate this, let $n_{\text{red}}$ and $n_{\text{blue}}$ denote the total number of students in two groups. For any student $r$ belonging to the red group, $S_r=n_{\text{red}}-1$, and for a student $b$ in the blue group $S_b=n_{\text{blue}}-1$, while for both groups $N_r=N_b=n_{\text{red}}+n_{\text{blue}}-1$. Therefore, we get that
\[
p_r + p_b = \frac{n_{\text{blue}}}{n_{\text{red}}+n_{\text{blue}}-1}+\frac{n_{\text{red}}}{n_{\text{red}}+n_{\text{blue}}-1}=\frac{n_{\text{blue}}+n_{\text{red}}}{n_{\text{red}}+n_{\text{blue}}-1}>1.
\]
This distinction is important in our small-scale networks, as the exclusion of a single individual can significantly shift the relative proportions of available peers. 

Furthermore, notice that the discrete patterns visible in the distribution of observed homophily (such as the prominent horizontal and vertical clusters of points at $0.5$ in Figure \ref{fig:observed_homophily}) are a direct mathematical consequence of working with small, finite networks. In finite classrooms, some proportions occur much more often than others, and $0.5$ is particularly common because it arises whenever the relevant counts split evenly. By contrast, values such as $0.52$ require a much more specific numerator--denominator combination and are therefore observed less frequently. The visible lines in the plots are thus a mechanical consequence of working with fractional counts in small networks, not evidence of any special behavioural threshold at $0.5$.

\subsection*{Modelling preference as an urn problem}\label{sec:wallenius_model}

Our goal is to distinguish observed homophily from underlying preference. In classroom networks, this distinction is crucial because the pool of available alters is both finite and highly asymmetric: for many students, especially minorities, the number of same-group and out-group peers differs sharply, and each nomination changes the composition of the remaining pool. A model that treats each tie as an independent Bernoulli draw therefore conflates preference with opportunity and becomes particularly problematic when students nominate a non-trivial share of their classmates. For this reason, we model nominations as a sequence of biased draws \emph{without replacement} using the Wallenius non-central hypergeometric distribution \cite{wallenius1964biased,fog2008calculation}. This distribution has been used in other settings where selection occurs from a finite pool under differential weights, including network modelling \cite{casiraghi2021configuration}, fairness-aware ranking \cite{cartier2025hyperfa}, and portfolio analysis \cite{bolshakov2020manager}.

We assume that same-group alters have weight $\omega_i > 0$ and out-group alters weight $1$, so that $\omega_i$ captures the student's relative preference for same-group over out-group peers. The resulting sampling model is
\[
s_i \mid S_i,N_i,k_i,\omega_i \sim \mathrm{WalleniusNCH}(S_i,N_i,k_i,\omega_i).
\]
Under this formulation, $\omega_i = 1$ corresponds to neutral mixing conditional on the available pool, $\omega_i > 1$ indicates same-group preference, and $\omega_i < 1$ indicates out-group preference. For ease of interpretation, we map the weight parameter onto the unit interval via
\[
q_i = \frac{\omega_i}{1+\omega_i},
\]
which can be read as the probability of choosing an in-group peer in a balanced one-versus-one comparison. On this scale, $q_i = 0.5$ denotes no preference, values above $0.5$ indicate homophily, and values below $0.5$ indicate heterophily.

The Wallenius probability mass function for observing $s_i$ same-group nominations is
\[
\Pr(s_i \mid S_i,N_i,k_i,\omega_i)
= \binom{S_i}{s_i}\binom{N_i-S_i}{k_i-s_i}
\int_0^1 \left(1-t^{\omega_i / d_i}\right)^{s_i}
\left(1-t^{1 / d_i}\right)^{k_i-s_i}\,dt,
\]
where
\[
d_i = \omega_i (S_i-s_i) + (N_i-S_i) - (k_i-s_i).
\]
This likelihood explicitly accounts for the depletion of the nomination pool as choices are made: when a student has already selected many same-group peers, the probability of another same-group nomination falls mechanically because fewer such alters remain available. The model therefore separates two forces that are confounded in raw homophily measures: a behavioural component encoded by $\omega_i$ and a structural component encoded by $(S_i,N_i,k_i)$.

This choice is substantively important for the present application. First, classrooms are small enough that finite-pool effects are not negligible; in some cases, students nominate a large fraction of all eligible classmates. Second, the model yields an individual-level preference parameter that can be embedded naturally in the multilevel model introduced next, allowing us to test whether preference itself varies systematically with classroom composition rather than merely reflecting changing opportunity structures. Although the Wallenius likelihood involves a one-dimensional integral, reliable numerical evaluation is robust with current software implementations \cite{fog2008calculation}.

\subsection*{The multilevel model}

To test whether preference varies systematically with classroom composition, we embed the Wallenius likelihood in a multilevel regression model. For student $i$ in group $g_i$ and classroom $c_i$, we model the log-odds preference parameter $\eta_i=\log(\omega_i)=\text{logit}(q_i)$ as
\[
s_i \mid \eta_i \sim \mathrm{WalleniusNCH}(S_i,N_i,k_i,\omega_i = e^{\eta_i}),
\qquad
\eta_i = \alpha_{g_i} + \beta_{g_i}(p_{c_i}-0.5) + u_{c_i} + v_i,
\]
with classroom- and student-level random effects $u_{c_i}\sim \mathcal{N}(0,\sigma_{\text{class}}^2)$ and $v_i \sim \mathcal{N}(0,\sigma_{\text{student}}^2)$. The intercept $\alpha_{g_i}$ captures baseline same-group preference in a compositionally balanced classroom, whereas $\beta_{g_i}$ captures how preference changes as the out-group share increases. Notice that we centre the out-group proportion $p_{c_i}$ by subtracting $0.5$, ensuring that the intercept $\alpha_{g_i}$ represents the baseline same-group preference in a compositionally balanced classroom where both groups are of equal size. Under structural opportunity theory, $\beta_g=0$; contact and conflict theory instead predict directional departures from zero, with the expected sign depending on the type of tie (friendship or rejection) and social dimension. 

Because the Wallenius probability mass function already requires numerical integration, and the hierarchical specification adds classroom- and student-level random effects, the marginal likelihood involves three nested integrals and has no closed form. We therefore approximate the random-effects integrals using Gauss--Hermite quadrature, a standard likelihood-based approach when exact integration is infeasible \cite{golub1969calculation,liu1994note,pan2003gauss}. Full computational details are reported in Section \ref{SI_sec:multilevel} of the Supplementary Information.

\subsection*{The Bayesian hypothesis testing framework}

Our inferential target is the composition effect $\beta_g$ for each social dimension, tie type, and focal group. We compare three models that map directly onto the competing theories: a point-null model $H_0$ corresponding to structural opportunity theory ($\beta_g=0$), and two directional alternatives corresponding to contact and conflict theory. Because the sign associated with contact versus conflict depends on tie valence, we report posterior model probabilities for the null, the contact-consistent direction, and the conflict-consistent direction.

Since a full Bayesian version of the hierarchical Wallenius model is computationally impractical, we approximate the likelihood for the focal coefficient using the normal-approximation approach of Barto\v{s} and Wagenmakers, based on the maximum-likelihood estimate $\hat\beta_g$ and its Wald standard error \cite{bartovs2023general}. We then combine this approximate likelihood with a spike-and-slab prior \cite{mitchell1988bayesian}. The spike is a point mass at zero and represents the exact null hypothesis, whereas the slab is a continuous prior over non-zero effects. By splitting the slab into its positive and negative halves and assigning equal prior probability to all three models, we obtain posterior model probabilities that can be interpreted directly as the relative evidence for structural opportunity, contact, and conflict theory \cite{mitchell1988bayesian,klugkist2010bayesian}. Under equal prior model weights, Bayes factors between any two theories reduce to ratios of their posterior model probabilities. Full details are reported in Section \ref{SI_sec:bayesian} of the Supplementary Information.

\subsection*{Comparison with other models}

We also compared the Wallenius specification to two alternative models that were feasible to estimate with the a full Bayesian hierarchical model \cite{mcglothlin2018bayesian}: a binomial approximation (sampling with replacement, equivalent to the formulation in \cite{altenburger2018monophily}) and Fisher's non-central hypergeometric model \cite{fog2008sampling}. Full results are reported in Section \ref{SI_sec:model-comparison} of the Supplementary Information. Across all three specifications, the evidence is broadly consistent in one important respect: we do not find meaningful support for contact theory. The main difference concerns conflict theory for gender in friendship ties. The Wallenius model yields the strongest support for a conflict-consistent composition effect, whereas the binomial model attenuates this pattern and does not reproduce it with the same strength. This difference is substantively informative. Because the binomial model treats nominations as conditionally independent draws, it does not account for the depletion of the available pool as students make many nominations. The stronger conflict-consistent signal under Wallenius therefore appears to be tied precisely to the fact that the finite opportunity set is consumed as ties are formed.

The Fisher model gives results that are qualitatively closer to Wallenius than the binomial approximation does, which is reassuring from the standpoint of robustness to alternative without-replacement formulations. We nevertheless do not use Fisher as the main model because its weight parameter has a different interpretation from the Wallenius weight. In Fisher's model, the parameter is inherited from a conditional odds-ratio construction, whereas in Wallenius it is a sequential selection weight \cite{fog2008calculation}. As a consequence, the Fisher model does not yield the same direct interpretation of our individual preference parameter.

These comparisons also help situate our approach relative to broader network frameworks, such as stochastic block models (SBMs, \cite{holland1983stochastic}) and exponential random graph models (ERGMs, \cite{robins2007introduction}). Whilst standard SBMs and many ERGM specifications typically summarise assortative structure through global parameters, our inferential target is an individual-level preference parameter whose dependence on classroom composition can itself be modelled. In this sense, these global models are designed to answer a fundamentally different set of questions regarding network architecture.


\printbibliography

@book{allport1954nature,
  title={The Nature of Prejudice},
  author={Allport, Gordon W},
  publisher={Addison-Wesley},
  year={1954}
}

@article{sajjadi2024unveiling,
  title={Unveiling homophily beyond the pool of opportunities},
  author={Sajjadi, Sina and Martin-Gutierrez, Samuel and Karimi, Fariba},
  journal={arXiv preprint arXiv:2401.13642},
  year={2024}
}

@article{bellmore2007influence,
  title={The influence of classroom ethnic composition on same-and other-ethnicity peer nominations in middle school},
  author={Bellmore, Amy D and Nishina, Adrienne and Witkow, Melissa R and Graham, Sandra and Juvonen, Jaana},
  journal={Social development},
  volume={16},
  number={4},
  pages={720--740},
  year={2007},
  publisher={Wiley Online Library}
}

@article{blumer1958race,
  title={Race prejudice as a sense of group position},
  author={Blumer, Herbert},
  journal={Pacific sociological review},
  volume={1},
  number={1},
  pages={3--7},
  year={1958},
  publisher={SAGE Publications Sage CA: Los Angeles, CA}
}

@article{boda2023ethnic,
  title={Ethnic diversity fosters the social integration of refugee students},
  author={Boda, Zs{\'o}fia and Lorenz, Georg and Jansen, Malte and Stanat, Petra and Edele, Aileen},
  journal={Nature Human Behaviour},
  volume={7},
  number={6},
  pages={881--891},
  year={2023},
  publisher={Nature Publishing Group UK London}
}

@article{cikara2022hate,
  author  = {Cikara, Mina and Fouka, Vasiliki and Tabellini, Marco},
  title   = {Hate crime towards minoritized groups increases as they increase in size-based rank},
  journal = {Nature Human Behaviour},
  year    = {2022},
  volume  = {6},
  number  = {11},
  pages   = {1537--1544},
  doi     = {10.1038/s41562-022-01416-5},
}

@article{hajdu2021ethnic,
  title={Ethnic segregation and inter-ethnic relationships in Hungarian schools},
  author={Hajdu, Tam{\'a}s and Kertesi, G{\'a}bor and K{\'e}zdi, G{\'a}bor},
  journal={On education. Journal for research and debate},
  volume={4},
  number={11},
  year={2021}
}

@article{hjalmarsson2023not,
  title={Not next to you: Peer rejection, sociodemographic characteristics and the moderating effects of classroom composition},
  author={Hjalmarsson, Simon and Fallesen, Peter and Plenty, Stephanie},
  journal={Journal of youth and adolescence},
  volume={52},
  number={6},
  pages={1191--1205},
  year={2023},
  publisher={Springer}
}

@article{joyner2000school,
  title={School racial composition and adolescent racial homophily},
  author={Joyner, Kara and Kao, Grace},
  journal={Social science quarterly},
  pages={810--825},
  year={2000},
  publisher={JSTOR}
}

@article{kawabata2011significance,
  title={The significance of cross-racial/ethnic friendships: associations with peer victimization, peer support, sociometric status, and classroom diversity.},
  author={Kawabata, Yoshito and Crick, Nicki R},
  journal={Developmental Psychology},
  volume={47},
  number={6},
  pages={1763},
  year={2011},
  publisher={American Psychological Association}
}

@article{lintner2023ukrainian,
  title={Ukrainian refugees struggling to integrate into Czech school social networks},
  author={Lintner, Tom{\'a}{\v{s}} and Divi{\'a}k, Tom{\'a}{\v{s}} and {\v{S}}e{\v{d}}ov{\'a}, Kl{\'a}ra and Hlado, Petr},
  journal={Humanities and Social Sciences Communications},
  volume={10},
  number={1},
  pages={1--11},
  year={2023},
  publisher={Palgrave}
}

@article{quillian2003beyond,
  title={Beyond black and white: The present and future of multiracial friendship segregation},
  author={Quillian, Lincoln and Campbell, Mary E},
  journal={American Sociological Review},
  volume={68},
  number={4},
  pages={540--566},
  year={2003},
  publisher={Sage Publications Sage CA: Los Angeles, CA}
}

@article{smith2016ethnic,
  title={Ethnic composition and friendship segregation: Differential effects for adolescent natives and immigrants},
  author={Smith, Sanne and Van Tubergen, Frank and Maas, Ineke and McFarland, Daniel A},
  journal={American Journal of Sociology},
  volume={121},
  number={4},
  pages={1223--1272},
  year={2016},
  publisher={University of Chicago Press Chicago, IL}
}

@article{mcpherson2001birds,
  title={Birds of a feather: Homophily in social networks},
  author={McPherson, Miller and Smith-Lovin, Lynn and Cook, James M},
  journal={Annual Review of Sociology},
  volume={27},
  pages={415--444},
  year={2001}
}

@book{blau1977inequality,
  title={Inequality and heterogeneity: A primitive theory of social structure},
  author={Blau, Peter M},
  publisher={Free Press},
  year={1977}
}

@article{pettigrew1998intergroup,
  title={Intergroup contact theory},
  author={Pettigrew, Thomas F},
  journal={Annual Review of Psychology},
  volume={49},
  pages={65--85},
  year={1998}
}

@article{pettigrew2011recent,
  title={Recent advances in intergroup contact theory},
  author={Pettigrew, Thomas F and Tropp, Linda R and Wagner, Ulrich and Christ, Oliver},
  journal={International Journal of Intercultural Relations},
  volume={35},
  number={3},
  pages={271--280},
  year={2011}
}

@article{hewstone2014contact,
  title={Intergroup contact and intergroup conflict},
  author={Hewstone, Miles and Lolliot, Sophie and Swart, Hermann and Myers, Eleanor and Voci, Alberto and Al Ramiah, Ananthi and Cairns, Ed},
  journal={Peace and Conflict: Journal of Peace Psychology},
  volume={20},
  number={1},
  pages={39--53},
  year={2014}
}

@article{chan2025promoting,
  title={Promoting friendship diversity in ethnically diverse schools: The role of school climate},
  author={Chan, Melissa and Benner, Aprile D},
  journal={Journal of Youth and Adolescence},
  year={2025}
}

@article{witkow2025perceptions,
  title={Perceptions of diversity and intergroup friendships in secondary schools},
  author={Witkow, Melissa R and Bellmore, Amy and Graham, Sandra},
  journal={Developmental Psychology},
  year={2025}
}

@article{mckeown2025understanding,
  title={Understanding intergroup contact in educational contexts: Opportunities and limits},
  author={McKeown, Shelley and Dixon, John and others},
  journal={Social and Personality Psychology Compass},
  year={2025}
}

@article{boller2025friendship,
  title={Attribute consolidation and friendship segregation in classrooms},
  author={B{\"o}ller, Sarah and Kruse, Hanno},
  journal={Social Networks},
  year={2025}
}

@article{meleady2026longitudinal,
  title={A longitudinal multilevel analysis of contextual opportunities and cross-ethnic friendships},
  author={Meleady, Rose and Seger, Charles and Abrams, Dominic},
  journal={Journal of Personality and Social Psychology},
  year={2026}
}

@article{kretschmer2025strong,
  title={Homophily in strong versus weak social ties across adolescence},
  author={Kretschmer, Tina and others},
  journal={Social Forces},
  year={2025}
}

@article{hovestadt2025similarity,
  title={Similarity in socioeconomic background and peer influence in educational trajectories},
  author={Hovestadt, Lisa and others},
  journal={Journal of Adolescence},
  year={2025}
}

@misc{tabery2024syri,
  title = {SYRI – výzkum na školách},
  author = {Tabery, Paulína and Čadová, Naděžda and Vinopal, Jiří and Ďurďovič, Martin and Pilnáček, Matouš and Červenka, Jan and Tuček, Milan and Vitíková, Jana and Hájková, Helena and Plačková, Marie and Spurný, Martin and Kyselá, Monika and Weikertová, Štěpánka and Ježková, Karolína and Novotná, Lucie and Lintner, Tomáš and Šeďová, Klára and Sedláček, Martin and Greger, David},
  year = {2024},
  doi = {10.14473/CSDA/I17GF8},
  note = {CSDA, Version 2.0}
}

@article{lintner2025affective,
  title = {How affective relationships and classroom norms shape perceptions of aggressor, victim, and defender roles},
  author = {Lintner, Tomáš and Klocek, Adam and Ropovik, Ivan and Kollerová, Lenka},
  journal = {Aggressive Behavior},
  volume = {51},
  number = {1},
  pages = {e70020},
  year = {2025},
  doi = {10.1002/ab.70020}
}

@misc{lintner2021classroom,
  title = {Peer social networks in Czech lower-secondary classrooms},
  author = {Lintner, Tomáš and Šeďová, Klára},
  year = {2021},
  doi = {10.17632/5vzy6rykm7.1},
  note = {Mendeley Data}
}

@article{lintner2023collectivity,
  title = {Relational and interactional dynamic network data from Czech lower-secondary school students},
  author = {Lintner, Tomáš and Šeďová, Klára and Sedláček, Martin and Šalamounová, Zuzana and Švaříček, Roman and Malíková, Kateřina and Nekardová, Barbora},
  journal = {Data in Brief},
  volume = {51},
  pages = {109641},
  year = {2023},
  doi = {10.1016/j.dib.2023.109641}
}

@misc{konradova2012vietnamci,
  title = {Vietnamci v českých školách},
  author = {Konrádová, Iva},
  year = {2012},
  note = {Bachelor thesis, Technical University of Liberec},
  url = {https://theses.cz/id/rkoikn/}
}

@misc{vaskova2019participace,
  title = {Vliv vrstevnické skupiny na participaci žáků na výukové komunikaci},
  author = {Vašková, Kristýna},
  year = {2019},
  note = {Bachelor thesis, Masaryk University},
  url = {https://is.muni.cz/th/y5xse/}
}

@article{evtushenko2021paradox,
  title={The paradox of second-order homophily in networks},
  author={Evtushenko, Anna and Kleinberg, Jon},
  journal={Scientific Reports},
  volume={11},
  number={1},
  pages={13360},
  year={2021},
  publisher={Nature Publishing Group UK London}
}

@article{peel2018multiscale,
  title={Multiscale mixing patterns in networks},
  author={Peel, Leto and Delvenne, Jean-Charles and Lambiotte, Renaud},
  journal={Proceedings of the National Academy of Sciences},
  volume={115},
  number={16},
  pages={4057--4062},
  year={2018},
  publisher={National Academy of Sciences}
}

@article{k2025homophily,
  title={Homophily within and across groups},
  author={K. Rizi, Abbas and Michielan, Riccardo and Stegehuis, Clara and Kivel{\"a}, Mikko},
  journal={Nature communications},
  year={2025},
  publisher={Nature Publishing Group UK London}
}

@article{currarini2009economic,
  title={An economic model of friendship: Homophily, minorities, and segregation},
  author={Currarini, Sergio and Jackson, Matthew O and Pin, Paolo},
  journal={Econometrica},
  volume={77},
  number={4},
  pages={1003--1045},
  year={2009},
  publisher={Wiley Online Library}
}

@book{maccoby1998two,
  title     = {The Two Sexes: Growing Up Apart, Coming Together},
  author    = {Maccoby, Eleanor E.},
  year      = {1998},
  publisher = {Harvard University Press},
  address   = {Cambridge, MA}
}

@book{thorne1993gender,
  title     = {Gender Play: Girls and Boys in School},
  author    = {Thorne, Barrie},
  year      = {1993},
  publisher = {Rutgers University Press},
  address   = {New Brunswick, NJ}
}

@article{martin2011patterns,
  title   = {Patterns of gender development},
  author  = {Martin, Carol Lynn and Ruble, Diane N.},
  journal = {Annual Review of Psychology},
  volume  = {61},
  pages   = {353--381},
  year    = {2010}
}

@incollection{hill1983menstrual,
  title     = {The intensification of gender-related role expectations during early adolescence},
  author    = {Hill, John P. and Lynch, Mary Ellen},
  booktitle = {Girls at Puberty: Biological and Psychosocial Perspectives},
  editor    = {Brooks-Gunn, Jeanne and Petersen, Anne C.},
  pages     = {201--228},
  year      = {1983},
  publisher = {Plenum Press},
  address   = {New York}
}

@article{galambos2009gender,
  title   = {Gender development in adolescence},
  author  = {Galambos, Nancy L. and Berenbaum, Sheri A. and McHale, Susan M.},
  journal = {Handbook of Adolescent Psychology},
  volume  = {1},
  pages   = {305--357},
  year    = {2009}
}

@article{mehta2009samegender,
  title   = {Same-gender versus cross-gender friendship conceptions: Similar or different?},
  author  = {Mehta, Clare M. and Strough, JoNell},
  journal = {Developmental Review},
  volume  = {29},
  number  = {3},
  pages   = {201--220},
  year    = {2009}
}

@article{cleveland1981lowess,
  title={LOWESS: A program for smoothing scatterplots by robust locally weighted regression},
  author={Cleveland, William S},
  journal={The American Statistician},
  volume={35},
  number={1},
  pages={54},
  year={1981},
  publisher={JSTOR}
}

@techreport{wallenius1964biased,
    author = {Wallenius, Kenneth Ted},
    title = {Biased sampling: the noncentral hypergeometric probability distribution},
    institution = {Stanford University},
    year = {1963}
}

@article{fog2008calculation,
  title={Calculation methods for Wallenius' noncentral hypergeometric distribution},
  author={Fog, Agner},
  journal={Communications in Statistics—Simulation and Computation{\textregistered}},
  volume={37},
  number={2},
  pages={258--273},
  year={2008},
  publisher={Taylor \& Francis}
}

@article{fog2008sampling,
  title={Sampling methods for Wallenius' and Fisher's noncentral hypergeometric distributions},
  author={Fog, Agner},
  journal={Communications in Statistics—Simulation and Computation{\textregistered}},
  volume={37},
  number={2},
  pages={241--257},
  year={2008},
  publisher={Taylor \& Francis}
}

@article{casiraghi2021configuration,
  title={Configuration models as an urn problem},
  author={Casiraghi, Giona and Nanumyan, Vahan},
  journal={Scientific reports},
  volume={11},
  number={1},
  pages={13416},
  year={2021},
  publisher={Nature Publishing Group UK London}
}

@inproceedings{cartier2025hyperfa,
  title={hyperFA* IR: A hypergeometric approach to fair rankings with finite candidate pool},
  author={Cartier van Dissel, Mauritz N and Martin-Gutierrez, Samuel and Esp{\'\i}n-Noboa, Lisette and Jaramillo, Ana Mar{\'\i}a and Karimi, Fariba},
  booktitle={Proceedings of the 2025 ACM Conference on Fairness, Accountability, and Transparency},
  pages={2112--2126},
  year={2025}
}

@article{bolshakov2020manager,
  title={Manager skill and portfolio size with respect to a benchmark},
  author={Bolshakov, Andrei and Chincarini, Ludwig B},
  journal={European Financial Management},
  volume={26},
  number={1},
  pages={176--197},
  year={2020},
  publisher={Wiley Online Library}
}

@article{bartovs2023general,
  title={A general approximation to nested Bayes factors with informed priors},
  author={Barto{\v{s}}, Franti{\v{s}}ek and Wagenmakers, Eric-Jan},
  journal={Stat},
  volume={12},
  number={1},
  pages={e600},
  year={2023},
  publisher={Wiley Online Library}
}

@article{liu1994note,
  title={A note on Gauss—Hermite quadrature},
  author={Liu, Qing and Pierce, Donald A},
  journal={Biometrika},
  volume={81},
  number={3},
  pages={624--629},
  year={1994},
  publisher={Oxford University Press}
}

@article{pinheiro2006efficient,
  title={Efficient Laplacian and adaptive Gaussian quadrature algorithms for multilevel generalized linear mixed models},
  author={Pinheiro, Jos{\'e} C and Chao, Edward C},
  journal={Journal of Computational and Graphical Statistics},
  volume={15},
  number={1},
  pages={58--81},
  year={2006},
  publisher={Taylor \& Francis}
}

@article{rabe2005maximum,
  title={Maximum likelihood estimation of limited and discrete dependent variable models with nested random effects},
  author={Rabe-Hesketh, Sophia and Skrondal, Anders and Pickles, Andrew},
  journal={Journal of Econometrics},
  volume={128},
  number={2},
  pages={301--323},
  year={2005},
  publisher={Elsevier}
}

@article{klugkist2010bayesian,
  title={Bayesian evaluation of inequality and equality constrained hypotheses for contingency tables.},
  author={Klugkist, Irene and Laudy, Olav and Hoijtink, Herbert},
  journal={Psychological methods},
  volume={15},
  number={3},
  pages={281},
  year={2010},
  publisher={American Psychological Association}
}

@article{mitchell1988bayesian,
  title={Bayesian variable selection in linear regression},
  author={Mitchell, Toby J and Beauchamp, John J},
  journal={Journal of the american statistical association},
  volume={83},
  number={404},
  pages={1023--1032},
  year={1988},
  publisher={Taylor \& Francis}
}

@book{weisberg2005applied,
  title={Applied linear regression},
  author={Weisberg, Sanford},
  volume={528},
  year={2005},
  publisher={John Wiley \& Sons}
}

@article{lee2019homophily,
  title={Homophily and minority-group size explain perception biases in social networks},
  author={Lee, Eun and Karimi, Fariba and Wagner, Claudia and Jo, Hang-Hyun and Strohmaier, Markus and Galesic, Mirta},
  journal={Nature Human Behaviour},
  volume={3},
  number={10},
  pages={1078--1087},
  year={2019},
  publisher={Nature Publishing Group UK London}
}

@article{altenburger2018monophily,
  title={Monophily in social networks introduces similarity among friends-of-friends},
  author={Altenburger, Kristen M and Ugander, Johan},
  journal={Nature human behaviour},
  volume={2},
  number={4},
  pages={284--290},
  year={2018},
  publisher={Nature Publishing Group UK London}
}

@article{martin2025intersectional,
  title={Intersectional inequalities in social ties},
  author={Martin-Gutierrez, Samuel and Cartier van Dissel, Mauritz N and Karimi, Fariba},
  journal={Science Advances},
  volume={11},
  number={45},
  pages={eadu9025},
  year={2025},
  publisher={American Association for the Advancement of Science}
}

@article{martin2026simple,
  title={A simple preference aggregation rule explains how multidimensional identities shape social networks},
  author={Martin-Gutierrez, Samuel and Cartier van Dissel, Mauritz N and Karimi, Fariba},
  journal={Communications Physics},
  year={2026},
  publisher={Nature Publishing Group UK London}
}

@article{byrd1995limited,
  title={A limited memory algorithm for bound constrained optimization},
  author={Byrd, Richard H and Lu, Peihuang and Nocedal, Jorge and Zhu, Ciyou},
  journal={SIAM Journal on scientific computing},
  volume={16},
  number={5},
  pages={1190--1208},
  year={1995},
  publisher={SIAM}
}

@ARTICLE{virtanen2020scipy,
  author  = {Virtanen, Pauli and Gommers, Ralf and Oliphant, Travis E. and
            Haberland, Matt and Reddy, Tyler and Cournapeau, David and
            Burovski, Evgeni and Peterson, Pearu and Weckesser, Warren and
            Bright, Jonathan and {van der Walt}, St{\'e}fan J. and
            Brett, Matthew and Wilson, Joshua and Millman, K. Jarrod and
            Mayorov, Nikolay and Nelson, Andrew R. J. and Jones, Eric and
            Kern, Robert and Larson, Eric and Carey, C J and
            Polat, {\.I}lhan and Feng, Yu and Moore, Eric W. and
            {VanderPlas}, Jake and Laxalde, Denis and Perktold, Josef and
            Cimrman, Robert and Henriksen, Ian and Quintero, E. A. and
            Harris, Charles R. and Archibald, Anne M. and
            Ribeiro, Ant{\^o}nio H. and Pedregosa, Fabian and
            {van Mulbregt}, Paul and {SciPy 1.0 Contributors}},
  title   = {{{SciPy} 1.0: Fundamental Algorithms for Scientific
            Computing in Python}},
  journal = {Nature Methods},
  year    = {2020},
  volume  = {17},
  pages   = {261--272},
  adsurl  = {https://rdcu.be/b08Wh},
  doi     = {10.1038/s41592-019-0686-2},
}

@article{carpenter2017stan,
  title={Stan: A probabilistic programming language},
  author={Carpenter, Bob and Gelman, Andrew and Hoffman, Matthew D and Lee, Daniel and Goodrich, Ben and Betancourt, Michael and Brubaker, Marcus and Guo, Jiqiang and Li, Peter and Riddell, Allen},
  journal={Journal of statistical software},
  volume={76},
  pages={1--32},
  year={2017}
}

@article{dickey1970weighted,
  title={The weighted likelihood ratio, sharp hypotheses about chances, the order of a Markov chain},
  author={Dickey, James M and Lientz, Bennet P},
  journal={The Annals of Mathematical Statistics},
  pages={214--226},
  year={1970},
  publisher={JSTOR}
}

@article{wagenmakers2010bayesian,
  title={Bayesian hypothesis testing for psychologists: A tutorial on the Savage--Dickey method},
  author={Wagenmakers, Eric-Jan and Lodewyckx, Tom and Kuriyal, Himanshu and Grasman, Raoul},
  journal={Cognitive psychology},
  volume={60},
  number={3},
  pages={158--189},
  year={2010},
  publisher={Elsevier}
}

@article{karimi2018homophily,
  title={Homophily influences ranking of minorities in social networks},
  author={Karimi, Fariba and G{\'e}nois, Mathieu and Wagner, Claudia and Singer, Philipp and Strohmaier, Markus},
  journal={Scientific reports},
  volume={8},
  number={1},
  pages={11077},
  year={2018},
  publisher={Nature Publishing Group UK London}
}

@article{holland1983stochastic,
  title={Stochastic blockmodels: First steps},
  author={Holland, Paul W and Laskey, Kathryn Blackmond and Leinhardt, Samuel},
  journal={Social networks},
  volume={5},
  number={2},
  pages={109--137},
  year={1983},
  publisher={Elsevier}
}

@article{robins2007introduction,
  title={An introduction to exponential random graph (p*) models for social networks},
  author={Robins, Garry and Pattison, Pip and Kalish, Yuval and Lusher, Dean},
  journal={Social networks},
  volume={29},
  number={2},
  pages={173--191},
  year={2007},
  publisher={Elsevier}
}

@article{mcglothlin2018bayesian,
  title={Bayesian hierarchical models},
  author={McGlothlin, Anna E and Viele, Kert},
  journal={Jama},
  volume={320},
  number={22},
  pages={2365--2366},
  year={2018}
}

@manual{fog2024urntheory,
  title = {Biased Urn Theory},
  author = {Fog, Agner},
  year = {2024},
  month = jun,
  note = {Vignette for the R package {BiasedUrn}},
  url = {https://archive.linux.duke.edu/cran/web/packages/BiasedUrn/vignettes/UrnTheory.pdf},
  urldate = {2026-05-04}
}

@article{golub1969calculation,
  title={Calculation of Gauss quadrature rules},
  author={Golub, Gene H and Welsch, John H},
  journal={Mathematics of computation},
  volume={23},
  number={106},
  pages={221--230},
  year={1969}
}

@article{pan2003gauss,
  title={Gauss-Hermite quadrature approximation for estimation in generalised linear mixed models},
  author={Pan, Jianxin and Thompson, Robin},
  journal={Computational statistics},
  volume={18},
  number={1},
  pages={57--78},
  year={2003},
  publisher={Springer}
}

\clearpage
\setcounter{section}{0}
\setcounter{subsection}{0}
\setcounter{equation}{0}
\setcounter{figure}{0}
\setcounter{table}{0}
\setcounter{page}{1}
\renewcommand{\thesection}{S\arabic{section}}
\renewcommand{\thesubsection}{S\arabic{section}.\arabic{subsection}}
\renewcommand{\theequation}{S\arabic{equation}}
\renewcommand{\thefigure}{S\arabic{figure}}
\renewcommand{\thetable}{S\arabic{table}}
\begin{center}
\textbf{\large Supplementary Information: Contact, Conflict, or Opportunity? Group Sizes and Intergroup Preference in Networks}
\end{center}

\section{Data description}\label{SI_sec:data_description}

All datasets were analysed cross-sectionally. Although several of the original studies employed longitudinal research designs, only the first available measurement wave was retained for the present analyses in order to ensure comparability in network timing, cohort exposure, and relational opportunity structures. 

Across datasets, affective peer relations were operationalised using sociometric nominations or ratings capturing positive and negative interpersonal orientations. To enable cross-study comparability, positive relational choices (e.g., liking, friendship, preferred collaboration partners) were harmonised and treated as indicators of \textit{friendship ties}, whereas negative relational choices (e.g., dislike, avoidance, unwillingness to share space) were harmonised and treated as indicators of \textit{rejection ties}. In datasets based on rating-scale sociometric instruments, thresholds distinguishing positive from neutral and negative evaluations were applied to reconstruct directed binary networks comparable to nomination-based measures.

\paragraph{PanelSYRI.}
The \textit{PanelSYRI} dataset \cite{tabery2024syri} provides a nationally stratified sample of sixth-grade classrooms across the Czech Republic. Data were collected longitudinally in 2023, 2024, and 2026; only the first wave is used here. Friendship ties were measured using an unlimited nomination item listing all classmates: \textit{``Write the names of the classmates you are friends with. You can write as many names as you want. The order of the names does not make any difference.''} Rejection ties were measured analogously with the item: \textit{``Write the names of the classmates with whom you would not like to share a desk.''} Both items allowed unrestricted nominations.

\paragraph{Defending Victims of Bullying.}
The \textit{Defending Victims of Bullying} dataset \cite{lintner2025affective} comprises seventh-grade classrooms from a convenience sample of Prague elementary schools. Twenty out of twenty-eight randomly contacted schools agreed to participate, yielding 39 classrooms. Data were collected at two time points within the 2015/2016 school year using paper-and-pencil questionnaires administered by trained research assistants; only the first wave is analysed here. Positive and negative affective relations were captured through unlimited relational nominations based on the questions \textit{``Who do you like most?''} and \textit{``Who do you like least?''} Directed binary ties were constructed to indicate the presence or absence of nominations.

\paragraph{Classroom Discourse.}
The \textit{Classroom Discourse} dataset \cite{lintner2021classroom} includes a non-probability sample of 435 ninth-grade students in 21 classrooms across 14 lower-secondary schools in the South Moravian Region. Data were collected in late 2017 using a standardised sociometric rating instrument adapted from Czech sociometric traditions. Students evaluated all classmates on a scale capturing likeability, neutrality, and antipathy. Directed friendship and rejection networks were reconstructed by dichotomising positive and negative evaluations.

\paragraph{Collectivity.}
The \textit{Collectivity} dataset \cite{lintner2023collectivity} comprises 276 sixth-grade students in twelve classrooms in the South Moravian Region. Although data were collected longitudinally at the beginning and end of the 2021/2022 school year, only the first measurement is used. Friendship ties were measured using a single unlimited nomination item worded identically to the PanelSYRI instrument. Questionnaires were administered in classroom settings by trained researchers.

\paragraph{Ukrainian Refugees.}
The \textit{Ukrainian Refugees} dataset \cite{lintner2023ukrainian} includes 266 students in grades five to nine from twelve classrooms in six Brno schools. Schools with relatively high enrolment of Ukrainian refugee students were purposively selected, and two mixed classrooms per school were randomly chosen. Data were collected in autumn 2022 using a bilingual Czech–Ukrainian sociometric questionnaire containing unlimited nomination items capturing friendship and rejection ties. Questionnaires were administered collectively during school lessons with researcher assistance.

\paragraph{Vietnamese in Czech Schools.}
The \textit{Vietnamese in Czech Schools} dataset \cite{konradova2012vietnamci} was collected in four elementary schools in the {\v C}esk{\'a} L{\'i}pa district. Classrooms were selected based on the presence of at least one Vietnamese student. Sociometric nominations were constrained to a maximum of three positive and three negative choices within each classroom. Positive and negative directed ties were derived from these limited-choice sociometric items.

\paragraph{Peer Group \& Participation.}
The \textit{Peer Group \& Participation} dataset \cite{vaskova2019participace} consists of two fifth-grade classrooms from two purposively selected schools in the Svitavy and Brno-venkov districts. Peer relations were measured through multiple open nomination items capturing preferred and non-preferred collaboration partners across several school situations (e.g., seating choice, school trips, best friends). Positive and negative ties were aggregated across items to reconstruct directed friendship and rejection networks.

\subsection{Number of data samples}\label{SI_sec:number_samples}

As mentioned in the main text, we applied a data pre-processing procedure. We started with $4,697$ students among $228$ classrooms, and the final sample sizes remaining after the application of these data cleaning and preprocessing criteria are summarised in Table \ref{tab:sample_sizes}. This table provides a detailed overview of the number of students and classrooms retained for each specific attribute-link combination and demographic subgroup.

\begin{table}[htbp]
\centering
\small
\begin{tabular}{lll rr}
\hline
Attribute & Link type & Group & $N$ students & $N$ classrooms \\
\hline

\multirow{6}{*}{Gender}
 & \multirow{3}{*}{Friendship}
 & Female & 1{,}818 & \multirow{2}{*}{ } \\
 &           & Male   & 1{,}836 &                     \\
 &           & Total  & 3{,}654 & 212                 \\
\cline{2-5}
 & \multirow{3}{*}{Rejection}
 & Female & 1{,}552 & \multirow{2}{*}{ } \\
 &           & Male   & 1{,}514 &                     \\
 &           & Total  & 3{,}066 & 198                 \\
\hline
\multirow{6}{*}{Ethnicity}
 & \multirow{3}{*}{Friendship}
 & Czech     & 669 & \multirow{2}{*}{ } \\
 &           & Non-Czech & 157 &                    \\
 &           & Total     & 826 & 45                 \\
 \cline{2-5}
 & \multirow{3}{*}{Rejection}
 & Czech     & 616 & \multirow{2}{*}{ } \\
 &           & Non-Czech & 137 &                    \\
 &           & Total     & 753 & 44                 \\
\hline
\multirow{6}{*}{SES}
 & \multirow{3}{*}{Friendship}
 & High SES & 1{,}044 & \multirow{2}{*}{ } \\
 &           & Low SES  & 1{,}505 &                      \\
 &           & Total    & 2{,}549 & 153                  \\
 \cline{2-5}
 & \multirow{3}{*}{Rejection}
 & High SES & 842     & \multirow{2}{*}{ } \\
 &           & Low SES  & 1{,}241 &                      \\
 &           & Total    & 2{,}083 & 142                  \\

\hline
\end{tabular}
\caption{Sample sizes after data cleaning.}
\label{tab:sample_sizes}
\end{table}

\subsection{Relationship between out-group proportion and out-degree for gender}\label{SI_sec:outgroup_outdegree}

In the Discussion of the main text, we note that one predicted consequence of the conflict-consistent behaviour observed in same-gender friendship preferences is that students belonging to the gender minority within a classroom may form fewer friendships overall than their majority-gender counterparts. Here we examine whether this pattern is present in our data.

\begin{figure}[htbp]
    \centering
    \includegraphics[width=0.6\linewidth]{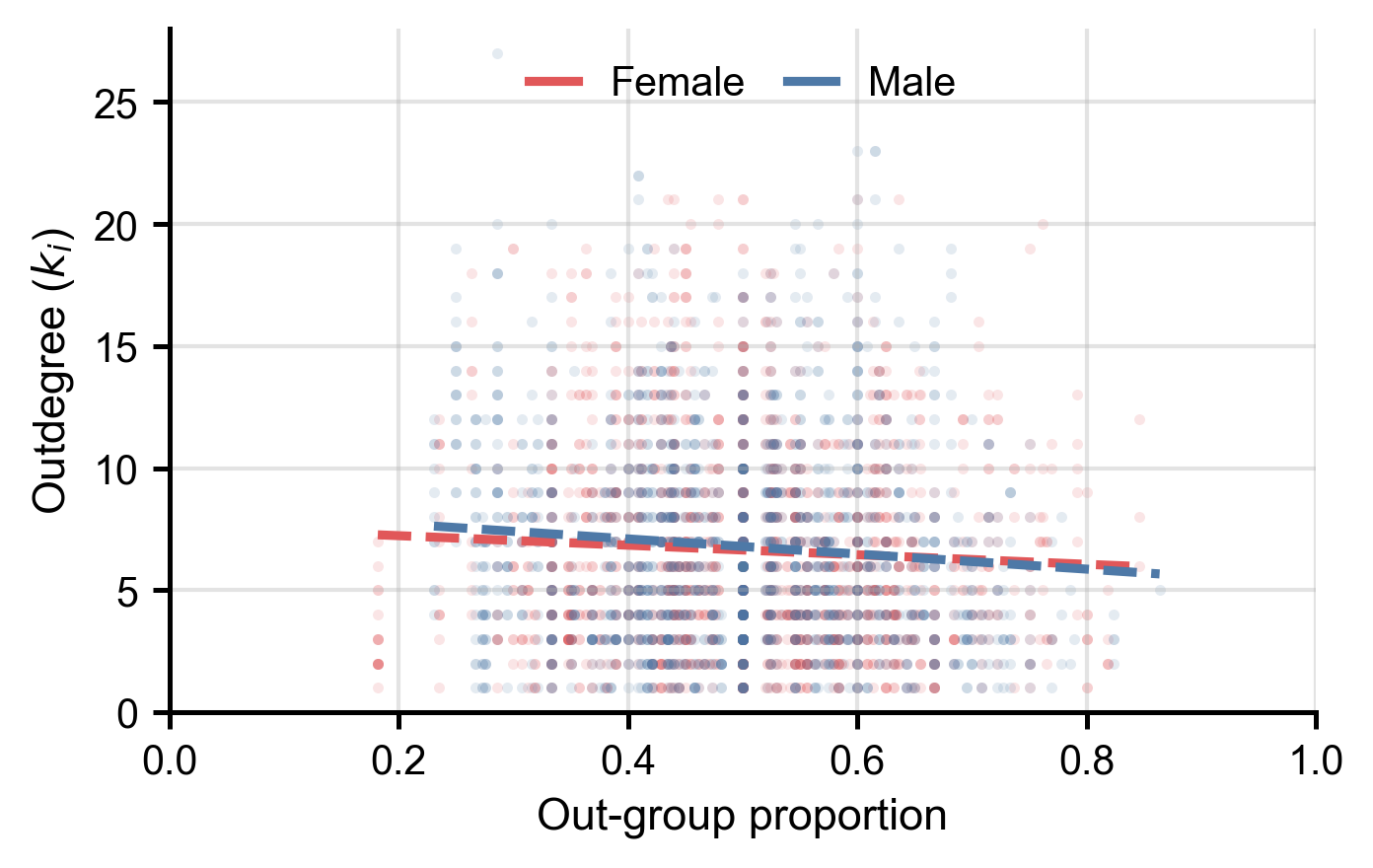}
    \caption{Relationship between classroom out-group proportion and observed outdegree. The figure shows that students in classrooms with a larger out-group share tend to make fewer nominations overall, with a steeper negative association for boys than for girls. }
    \label{fig:outgroup-outdegree}
\end{figure}

Figure~\ref{fig:outgroup-outdegree} and Table~\ref{tab:outgroup-outdegree} summarise the relationship between the classroom out-group proportion and observed outdegree ($k_i$), estimated separately for girls and boys. For both genders, a higher out-group proportion is associated with a lower outdegree, and the negative association is steeper for boys (slope$=-3.09$, p-value$<0.001$) than for girls (slope$=-1.95$, p-value$=0.016$). Both slopes are statistically significant, indicating that students in more gender-imbalanced classrooms—those in which their own gender is in the minority—nominate fewer peers overall.

This reduction in nominations could itself be a consequence of the shift in same-gender preference reported in Figure \ref{fig:preference} of the main text: as the available pool of same-gender peers shrinks, students may not fully substitute toward out-group nominations, leading to fewer ties overall. Disentangling the causal link of this interpretation is beyond the scope of the present analysis and is an important direction for future work.

\begin{table}[htbp]
\centering
\small
\begin{tabular}{lrrrr}
\hline
Group & Slope & Std. error & $p$-value & $n$ \\
\hline
Female & $-1.9542$ & $0.8115$ & $0.0161$ & $1818$ \\
Male   & $-3.0942$ & $0.8119$ & $0.0001$ & $1836$ \\
\hline
\end{tabular}
\caption{Linear association between classroom out-group proportion and observed outdegree by gender. Negative slopes indicate that students in classrooms with a larger out-group share tend to nominate fewer peers in total.}
\label{tab:outgroup-outdegree}
\end{table}

\section{Multilevel model: technical details}\label{SI_sec:multilevel}

The main text introduces the hierarchical Wallenius model. Here we give the technical details regarding its computations. To reintroduce the model, for student $i$ in focal group $g_i$ and classroom $c_i$, we observe $s_i$ same-group nominations out of $k_i$ total nominations, with $S_i$ available same-group alters out of $N_i$ eligible alters in total. The individual preference parameter is written on the log scale as
\[
\eta_i = \alpha_{g_i} + \beta_{g_i}(p_{c_i}-0.5) + u_{c_i} + v_i,
\qquad p_{c_i}=\frac{N_i-S_i}{N_i},
\]
with classroom and student random effects
\[
u_c \overset{\mathrm{iid}}{\sim} \mathcal{N}(0,\sigma_u^2),
\qquad
v_i \overset{\mathrm{iid}}{\sim} \mathcal{N}(0,\sigma_v^2).
\]
The Wallenius weight is $\omega_i=e^{\eta_i}$, so the sampling model is
\[
s_i\mid \eta_i \sim \mathrm{WalleniusNCH}(S_i,N_i,k_i,e^{\eta_i}).
\]
The intercept $\alpha_g$ captures baseline same-group preference in a compositionally balanced classroom, whereas $\beta_g$ captures how preference changes as the out-group share increases.

\subsection{Likelihood computation using Gauss--Hermite quadrature}\label{SI_sec:ghq}

In the methods, we mention that we use Gauss-Hermite quadrature to calculate the marginal likelihood for our multilevel model. The marginal likelihood integrates over both levels of Gaussian random effects. For classroom $c$, let $\mathcal{I}_c$ denote the set of students in that classroom. The classroom-specific contribution is
\[
L_c(\theta)
=
\int_{-\infty}^{\infty}
\left[\prod_{i\in \mathcal{I}_c}
\int_{-\infty}^{\infty}
 p\bigl(s_i\mid S_i,N_i,k_i,\eta_i(u_c,v_i;\theta)\bigr)
 \,\phi(v_i;0,\sigma_v^2)
 \,dv_i
\right]
\phi(u_c;0,\sigma_u^2)
\,du_c,
\]
where $\theta$ collects the fixed effects and variance parameters, $p(\cdot)$ is the Wallenius probability mass function, and $\phi(\cdot)$ is the probability density function of the normal distribution. The inner integral averages over the student-specific random effect conditional on the classroom effect, and the outer integral then averages over the shared classroom effect. The full marginal likelihood is then $L(\theta)=\prod_c L_c(\theta)$.

We approximate these Gaussian integrals by non-adaptive Gauss--Hermite quadrature \cite{golub1969calculation, pan2003gauss} with 15 nodes per random-effect dimension. Writing $\{x_m,w_m\}_{m=1}^{15}$ for the standard Gauss--Hermite nodes and weights, a generic one-dimensional Gaussian integral is approximated as
\[
\int f(z)\phi(z;0,\sigma^2)\,dz
\approx
\frac{1}{\sqrt{\pi}}\sum_{m=1}^{15} w_m \, f\!\left(\sqrt{2}\sigma x_m\right).
\]
Applying this rule to $u_c$ and to each student-level random effect yields a nested quadrature approximation to $L_c(\theta)$. Inside each quadrature evaluation, the Wallenius probability mass function still contains its own one-dimensional integral, which we evaluate numerically using Fog's implementation in the BiasedUrn package in R \cite{fog2008calculation, fog2024urntheory}.

We use \emph{non-adaptive} rather than adaptive Gauss--Hermite quadrature \cite{liu1994note,rabe2005maximum,pinheiro2006efficient} because the Wallenius likelihood is already expensive to evaluate, so repeatedly re-centering and re-scaling the quadrature nodes around classroom-specific posterior modes and local curvature would add a large computational burden at every optimiser step. In our setting, where the random-effects dimension is moderate but the inner likelihood is nonstandard, a fixed 15-point rule provides a compromise between accuracy and computational tractability. Adaptive quadrature can be more accurate when the conditional likelihood is strongly non-Gaussian, but here that additional accuracy would come at substantial computational cost and less stable likelihood evaluation.

\subsection{Maximum marginal likelihood estimation}\label{SI_sec:mle}

We estimate the parameter vector by maximising the approximate marginal log-likelihood
\[
\ell(\theta)=\sum_c \log L_c(\theta).
\]
Optimisation is carried out with the L-BFGS-B algorithm as implemented in \texttt{scipy.optimize.minimize} \cite{byrd1995limited,virtanen2020scipy}. L-BFGS-B is a limited-memory quasi-Newton method that scales well to moderately high-dimensional parameter vectors and permits simple box constraints on variance parameters when needed. 

We report the maximiser $\hat\theta$ together with Wald uncertainty based on the inverse observed Hessian at the optimum. The resulting standard error for the focal composition coefficient $\hat\beta_g$ is denoted $\widehat{\mathrm{se}}(\hat\beta_g)$. We chose this maximum-likelihood strategy over full Bayesian estimation because the Wallenius distribution is not natively available in current software for Bayesian hierarchical models (e.g. Stan \cite{carpenter2017stan}), and embedding a custom Bayesian implementation would require repeated numerical integration within posterior sampling, which is computationally prohibitive for the present sample size and model complexity.

\section{Bayesian Hypothesis Testing}\label{SI_sec:bayesian}

\subsection{Posterior from a normal approximation}\label{SI_sec:normal-approximation}

For theory testing, our inferential target is the scalar coefficient $\beta_g$. Rather than fitting a full Bayesian version of the hierarchical Wallenius model, we approximate the likelihood for $\beta_g$ using the standard large-sample normal approximation implied by the maximum-likelihood estimator \cite{bartovs2023general}:
\[
\hat\beta_g \mid \beta_g \;\dot\sim\; \mathcal{N}\!\bigl(\beta_g,\widehat{\mathrm{se}}(\hat\beta_g)^2\bigr).
\]
Equivalently, up to proportionality, the approximate likelihood for $\beta_g$ is
\[
\mathcal{L}(\beta_g)
\propto
\exp\left[-\frac{1}{2}
\frac{(\hat\beta_g-\beta_g)^2}{\widehat{\mathrm{se}}(\hat\beta_g)^2}
\right].
\]
This is not a full posterior for all model parameters. Instead, it is an approximate one-dimensional likelihood for the focal coefficient that uses only the point estimate and its uncertainty. Combining this approximate likelihood with a prior on $\beta_g$ yields an approximate posterior and Bayes factors as described by Barto\v{s} and Wagenmakers \cite{bartovs2023general}. The advantage is that we can perform explicit model comparison for the composition effect without re-estimating the entire hierarchical model in a fully Bayesian framework.

\subsection{Spike-and-slab posterior model probabilities}\label{SI_sec:spike-slab}

We compare three hypotheses,
\[
M_0:\beta_g=0,
\qquad
M_+:\beta_g>0,
\qquad
M_-:\beta_g<0,
\]
corresponding to structural opportunity, contact, and conflict theory. Which sign corresponds to contact or conflict theory depends on whether we are focusing on friendship (contact is positive) or rejection (conflict is positive) ties, as described in the main text.

To assign non-zero posterior probability to the exact null, we use a spike-and-slab prior \cite{mitchell1988bayesian}:
\[
p(\beta_g)
=
\frac{1}{3}\,\delta_0(\beta_g)
+
\frac{2}{3}\,\mathcal N(\beta_g;0,\sigma^2),
\]
where the Gaussian slab is split symmetrically into its positive and negative halves, so each of $M_0$, $M_+$, and $M_-$ receives prior probability $1/3$. For our analysis, we use $\sigma=4$. Let
\[
m_0 = \mathcal{L}(0),
\qquad
m_+ = \int_0^{\infty} \mathcal{L}(\beta)\,\mathcal N(\beta;0,\sigma^2)\,d\beta,
\qquad
m_- = \int_{-\infty}^{0} \mathcal{L}(\beta)\,\mathcal N(\beta;0,\sigma^2)\,d\beta.
\]
These are the marginal likelihood contributions for the three models. Because the prior model weights are equal, the posterior model probabilities are simply
\[
\mathrm{PMP}_0 = \frac{m_0}{m_0+m_++m_-},
\qquad
\mathrm{PMP}_+ = \frac{m_+}{m_0+m_++m_-},
\qquad
\mathrm{PMP}_- = \frac{m_-}{m_0+m_++m_-}.
\]
Bayes factors between any two theories are then ratios of the corresponding posterior model probabilities.

\subsection{Connection to the Savage--Dickey density ratio and Bayesian hierarchical models}\label{SI_sec:savage-dickey}

Our spike-and-slab construction is closely related to the Savage--Dickey density ratio, a classical identity for Bayes factors for point-null hypotheses under a continuous prior \cite{dickey1970weighted,wagenmakers2010bayesian,bartovs2023general}. Under a purely continuous Gaussian prior, the Savage--Dickey representation expresses the Bayes factor for the point null $M_0:\beta=0$ against the encompassing alternative ($M_e:\beta\neq0$) in terms of the ratio of posterior to prior density at zero. In the present notation and with the previously defined slab as the prior, this yields
\[
\mathrm{BF}^{\mathrm{SD}}_{0,e}
=
\frac{p(\beta=0\mid D)}{p(\beta=0)}
.
\]
Using Bayes' rule,
\[
p(\beta\mid D)
=
\frac{\mathcal{L}(\beta)\,\mathcal N(\beta;0,\sigma^2)}{\int_{-\infty}^{\infty} \mathcal{L}(t)\,\mathcal N(t;0,\sigma^2)\,dt},
\]
so evaluating at $\beta=0$ and dividing by $p(\beta=0)=\mathcal N(0;0,\sigma^2)$ yields
\[
\mathrm{BF}^{\mathrm{SD}}_{0,e}
=
\frac{\mathcal{L}(0)}{\int_{-\infty}^{\infty} \mathcal{L}(t)\,\mathcal N(t;0,\sigma^2)\,dt}
=
\frac{\mathcal{L}(0)}{m_+ + m_-}.
\]
But this is exactly the spike-and-slab Bayes factor for the null against the undirected non-null alternative,
\[
\mathrm{BF}_{0,\neq 0}
=
\frac{p(D\mid M_0)}{p(D\mid M_{\neq 0})}
=
\frac{\mathcal{L}(0)}{m_+ + m_-}
=
\mathrm{BF}^{\mathrm{SD}}_{0,e}.
\]
Hence the Savage--Dickey ratio computed from a continuous model gives the same null-versus-non-null evidence as our spike-and-slab construction. 

The crucial difference concerns posterior probabilities. Under a purely continuous prior, the posterior probability of the exact event $\beta=0$ is necessarily zero, because a continuous prior assigns zero probability mass to any single point. By contrast, under spike-and-slab the null receives explicit prior mass through the spike, so its posterior probability remains positive and directly interpretable. This distinction is precisely what makes the spike-and-slab formulation useful for our setting. It preserves the simplicity associated with the Savage--Dickey ratio, while also allowing us to assign posterior model probabilities to three competing theories: structural opportunity, contact, and conflict.

This is also useful in a fully Bayesian hierarchical implementation (we will discuss this in the next section), since in current Bayesian software such as Stan, one cannot include an actual point-mass spike because discrete parameters are not allowed. A continuous-prior fit still provides the Savage--Dickey Bayes factor, and therefore also $\mathrm{BF}_{0,\neq 0}$. With equal prior model weights, $\pi_0=\pi_+=\pi_-=1/3$, the posterior model probability for the null is
\[
\mathrm{PMP}_0
=
\frac{\mathrm{BF}_{0,\neq 0}}{\mathrm{BF}_{0,\neq 0}+2}.
\]
The remaining mass, $1-\mathrm{PMP}_0$, is then split between the two directional alternatives using the posterior sign probabilities under the continuous slab,
\[
P(\beta>0\mid D,M_{\neq 0}) = \frac{m_+}{m_+ + m_-},
\qquad
P(\beta<0\mid D,M_{\neq 0}) = \frac{m_-}{m_+ + m_-}.
\]
Therefore the three posterior model probabilities are recovered as
\[
\mathrm{PMP}_+
=
(1-\mathrm{PMP}_0)\frac{m_+}{m_+ + m_-},
\qquad
\mathrm{PMP}_-
=
(1-\mathrm{PMP}_0)\frac{m_-}{m_+ + m_-}.
\]
In a fully Bayesian continuous-prior fit, these two fractions can be estimated from the posterior proportions of positive and negative draws. Thus a fully continuous model provides all ingredients needed for the three-theory comparison: the Savage--Dickey ratio determines the null-versus-non-null evidence, and the posterior sign probabilities partition the non-null mass into the contact-consistent and conflict-consistent directions

\section{Comparison with binomial and Fisher models}\label{SI_sec:model-comparison}

To assess how strongly our substantive conclusions depend on the Wallenius sampling mechanism, we also fit two alternative hierarchical models for the same focal coefficient $\beta_g$. In both alternatives, full Bayesian estimation is computationally feasible, so the models were implemented hierarchically in a fully Bayesian way using Stan \cite{carpenter2017stan} rather than through the maximum likelihood strategy used for Wallenius. We briefly discuss the choice of priors for the Bayesian models in Section \ref{SI_sec:priors}. Regarding the computation of the Posterior Model Probabilities for the three theories, we use the formulas derived in the previous subsection \ref{SI_sec:savage-dickey}.

\subsection{Binomial approximation}\label{SI_sec:binomial-comparison}

\begin{figure}[h]
    \centering
    \includegraphics[width=0.6\linewidth,trim=0cm 0cm 0cm 0.18cm,clip]{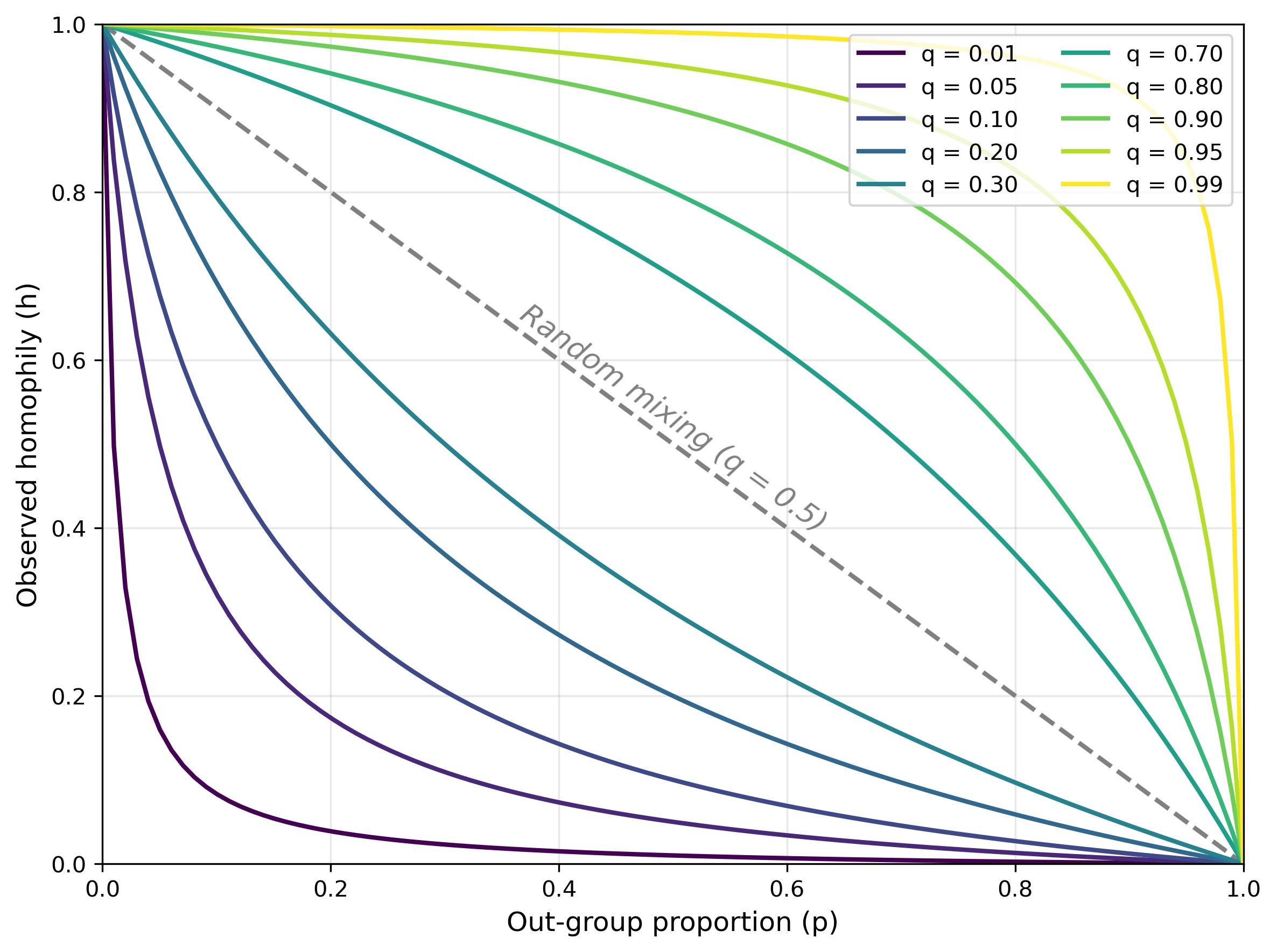}
    \caption{Observed homophily $h$ as a function of out-group proportion $p$ for different fixed latent preference values $q$ under the binomial approximation. The dashed line shows random mixing, $q=0.5$. Curves above the diagonal indicate homophilic preferences ($q>0.5$), and curves below it indicate heterophilic preferences ($q<0.5$). The figure illustrates that the same underlying preference can generate very different observed same-group nomination shares depending solely on classroom composition.}
    \label{fig:homophily_q}
\end{figure}

The first alternative corresponds to the binomial approximation used in \cite{altenburger2018monophily} and treats nominations as independent draws with a success probability that depends on both preference and classroom composition. Let $q_i$ denote the latent same-group preference, and let $p_i=\frac{N_i-S_i}{N_i}$ denote the share of available out-group alters. The induced probability of a same-group nomination, which corresponds to the first draw of our Wallenius model, is
\[
\pi_i = \frac{q_i (1-p_i)}{q_i (1-p_i) + (1-q_i)p_i}.
\]
Notice that if we apply the logit function, we can rewrite the relationship as
\[
\text{logit}(\pi_i) = \text{logit}(q_i) + \text{logit}(1-p_i).
\]
The observation model is then
\[
s_i \mid k_i,\pi_i \sim \mathrm{Binomial}(k_i,\pi_i).
\]
At the individual level, we defined observed homophily as $h_i=s_i/k_i$, that is, the observed fraction of same-group nominations, and we can observe that $h_i$ is the maximum-likelihood estimate of the induced success probability $\pi_i$. The latent preference parameter $q_i$ is different: it is the same-group choice probability that would apply in a balanced one-versus-one comparison, before accounting for the classroom composition term $p_i$. Solving the relationship above for $q_i$, the individual plug-in estimate of preference is obtained by replacing $\pi_i$ with its MLE $h_i$,
\[
\hat q_i = \frac{h_ip_i}{h_ip_i + (1-h_i)(1-p_i)} \qquad \& \qquad \text{logit}(\hat q_i) = \text{logit}(h_i) - \text{logit}(1-p_i).
\]
This makes the distinction between observed homophily and latent preference explicit for this approximation. When $p_i=0.5$, we have $\pi_i=q_i$, so observed homophily directly estimates preference. When $p_i\neq 0.5$, however, the same value of $h_i$ can correspond to different preference levels depending on the opportunity structure. In particular, a student in a strongly imbalanced classroom can display a low observed same-group share $h_i$ even when the implied preference $q_i$ remains clearly homophilic. Figure~\ref{fig:homophily_q} visualizes this relationship. For each fixed latent preference value $q$, the observed homophily $h$ declines mechanically as the out-group proportion $p$ increases, even when underlying preference is held constant. The dashed diagonal corresponds to random mixing, $q=0.5$, while curves above and below it represent homophilic and heterophilic preferences, respectively.

This model is a useful approximation because it preserves the basic distinction between preference and opportunity: the composition term $p_i$ shifts the success probability even when $q_i$ is held fixed. However, it treats nominations as conditionally independent and therefore ignores depletion of the available pool across successive nominations. In that sense, it is a with-replacement approximation to the without-replacement process represented by the Wallenius model.

Table~\ref{tab:bayes_binomial_asterisks} shows that the binomial results recover the broad qualitative picture but attenuate some composition effects. In particular, the binomial model still suggests a conflict-consistent pattern for female friendship nominations by gender, but the evidence is notably weaker than under Wallenius. For ethnicity and most SES rows, posterior mass remains concentrated on the null, again consistent with structural opportunity theory. The overall pattern therefore indicates that a simpler with-replacement approximation already captures part of the empirical signal, but it smooths over finite-pool effects and can pull extreme preference estimates back toward the center.

\begin{table}[htbp]
\centering
\small
\begin{tabular}{lll r rl rrr rr}
\hline
Attribute & Link type & Group
  & q($\alpha$)
  & $\beta$ & 95\% CI
  & PMP$_\text{contact}$ & PMP$_0$ & PMP$_\text{conflict}$
  & $\sigma_\text{class}$ & $\sigma_\text{student}$ \\
\hline

\multirow{4}{*}{Gender}
 & \multirow{2}{*}{Friendship}
 & Female  & $0.77^{*}$  & $0.92^{*}$  & $[0.02,\ 1.85]$  & $0.01$ & $0.37$ & $\mathbf{0.62}$  & \multirow{2}{*}{0.66} & \multirow{2}{*}{0.67} \\

 & 
 & Male  & $0.79^{*}$  & $0.30$  & $[-0.60,\ 1.18]$  & $0.06$ & $\mathbf{0.78}$ & $0.17$  &  &  \\

 & \multirow{2}{*}{Rejection}
 & Female  & $0.28^{*}$  & $-0.61$  & $[-1.73,\ 0.49]$  & $0.05$ & $\mathbf{0.66}$ & $0.30$  & \multirow{2}{*}{0.77} & \multirow{2}{*}{0.49} \\

 & 
 & Male  & $0.41^{*}$  & $-0.45$  & $[-1.59,\ 0.64]$  & $0.06$ & $\mathbf{0.72}$ & $0.22$  &  &  \\
\hline
\multirow{4}{*}{Ethnicity}
 & \multirow{2}{*}{Friendship}
 & Czech  & $0.60^{*}$  & $0.48$  & $[-1.07,\ 2.00]$  & $0.08$ & $\mathbf{0.68}$ & $0.24$  & \multirow{2}{*}{0.50} & \multirow{2}{*}{0.05} \\

 & 
 & Non-Czech  & $0.63^{*}$  & $-0.26$  & $[-1.85,\ 1.29]$  & $0.19$ & $\mathbf{0.71}$ & $0.11$  &  &  \\

 & \multirow{2}{*}{Rejection}
 & Czech  & $0.49$  & $-0.39$  & $[-2.73,\ 1.88]$  & $0.14$ & $\mathbf{0.62}$ & $0.24$  & \multirow{2}{*}{0.59} & \multirow{2}{*}{0.11} \\

 & 
 & Non-Czech  & $0.41$  & $-0.16$  & $[-2.38,\ 1.99]$  & $0.16$ & $\mathbf{0.65}$ & $0.20$  &  &  \\
\hline
\multirow{4}{*}{SES}
 & \multirow{2}{*}{Friendship}
 & High SES  & $0.51^{*}$  & $0.28$  & $[-0.07,\ 0.62]$  & $0.01$ & $\mathbf{0.77}$ & $0.22$  & \multirow{2}{*}{0.06} & \multirow{2}{*}{0.02} \\

 & 
 & Low SES  & $0.50$  & $0.36^{*}$  & $[0.02,\ {0.68}]$  & $0.01$ & $\mathbf{0.58}$ & $0.41$  &  &  \\

 & \multirow{2}{*}{Rejection}
 & High SES  & $0.48$  & $-0.48$  & $[-1.23,\ 0.26]$  & $0.03$ & $\mathbf{0.70}$ & $0.27$  & \multirow{2}{*}{0.38} & \multirow{2}{*}{0.03} \\

 & 
 & Low SES  & $0.53^{*}$  & $-0.02$  & $[-0.72,\ 0.69]$  & $0.07$ & $\mathbf{0.85}$ & $0.08$  &  &  \\

\hline
\end{tabular}
\caption{Bayesian binomial model results. $q(\alpha)$ denotes the baseline same-group preference probability at equal in-group and out-group classroom composition. $\beta$ is the posterior median composition effect, with 95\% credible interval shown in brackets. PMP$_0=1/(1+2\cdot\text{BF}_{10}(\beta))$ gives the posterior model probability of the null under equal prior model weights, whereas PMP$_\text{contact}$ and PMP$_\text{conflict}$ partition the remaining posterior mass between the two directional alternatives. Boldface marks the highest-probability direction in each row. $\sigma_\text{class}$ and $\sigma_\text{student}$ are posterior medians of the classroom-group and student-level standard deviations. Superscript significance marks q($\alpha$) when the 90\% credible interval excludes zero and marks $\beta$ according to posterior sign probability: $^*{>}0.95$, $^{**}{>}0.99$, $^{***}{>}0.999$.
}
\label{tab:bayes_binomial_asterisks}
\end{table}

\subsection{Fisher non-central hypergeometric model}\label{SI_sec:fisher-comparison}

The second alternative uses Fisher's non-central hypergeometric distribution \cite{fog2008sampling}. Like Wallenius, it is a without-replacement model for counts of same-group nominations. This distribution corresponds to the conditional distribution defined in \cite{altenburger2018monophily} if the with-replacement approximation was not used. Conditional on a fixed number of draws $k_i$, a fixed number of available in-group and out-group alters, and a bias parameter, it can be interpreted as the distribution induced by a dyad-independent choice model after conditioning on the sender's out-degree. Notably, without the student-level random effects, a classroom-level model using the Fisher distribution corresponds to an SBM \cite{holland1983stochastic} with fixed out-degree.

However, its parameter is not equivalent to the Wallenius weight. Fog shows that Wallenius' distribution is generated by \emph{sequential} without-replacement sampling, where the probability of each draw depends on the weights of the items remaining in the urn, whereas Fisher's distribution is obtained by conditioning \emph{independent} weighted draws on the final total number selected \cite{fog2008sampling,fog2024urntheory}. These are different stochastic mechanisms, and Fog further notes that the two distributions are only approximately similar when odds ratios are near 1 and the number of draws is small relative to the population size; otherwise they can differ substantially \cite{fog2024urntheory}. Therefore, a value such as $\omega=2$ does not carry the same interpretation in the two models and need not imply the same expected same-group nomination pattern. For the same reason, Fisher's model does not yield a direct probability-scale preference parameter $q$ analogous to the Wallenius model or the binomial approximation. In the binomial model we can separate preference and opportunity through the identity $\operatorname{logit}(\pi_i)=\operatorname{logit}(q_i)+\operatorname{logit}(p_i)$, but Fisher's non-central hypergeometric model is defined as a conditional distribution for the final allocation rather than through such a probability decomposition \cite{fog2024urntheory}. 


Table~\ref{tab:bayes_fisher_asterisks} yields conclusions that are directionally similar to the Wallenius model in the most prominent cases. In particular, the strongest departure from the null again appears for female friendship nominations by gender, where the posterior mass is concentrated on the conflict-consistent direction. At the same time, several other rows are more weakly identified, and the magnitude of the baseline preference parameters is not directly comparable to the Wallenius estimates because the underlying weight parameter has a different interpretation. The main takeaway is therefore not that Fisher reproduces Wallenius numerically, but that the central qualitative result---strongest composition dependence for gender friendship, weaker or null effects for ethnicity and most SES contrasts---is robust to an alternative without-replacement specification.

\begin{table}[htbp]
\centering
\small
\begin{tabular}{lll r rl rrr rr}
\hline
Attribute & Link type & Group
  & q($\alpha$)
  & $\beta$ & 95\% CI
  & PMP$_\text{contact}$ & PMP$_0$ & PMP$_\text{conflict}$
  & $\sigma_\text{class}$ & $\sigma_\text{student}$ \\
\hline

\multirow{4}{*}{Gender}
 & \multirow{2}{*}{Friendship}
 & Female  & $0.89^{*}$  & $2.59^{***}$  & $[1.11,\ {4.08}]$  & $<0.01$ & $0.01$ & $\mathbf{0.99}$  & \multirow{2}{*}{1.09} & \multirow{2}{*}{1.30} \\

 & 
 & Male  & $0.91^{*}$  & $1.17$  & $[-0.31,\ 2.65]$  & $0.03$ & $0.46$ & $\mathbf{0.51}$  &  &  \\

 & \multirow{2}{*}{Rejection}
 & Female  & $0.21^{*}$  & $-1.14$  & $[-2.68,\ 0.37]$  & $0.04$ & $0.46$ & $\mathbf{0.50}$  & \multirow{2}{*}{1.09} & \multirow{2}{*}{1.08} \\

 & 
 & Male  & $0.37^{*}$  & $-0.78$  & $[-2.31,\ 0.74]$  & $0.06$ & $\mathbf{0.61}$ & $0.33$  &  &  \\
\hline
\multirow{4}{*}{Ethnicity}
 & \multirow{2}{*}{Friendship}
 & Czech  & $0.68^{*}$  & $0.70$  & $[-2.23,\ 3.60]$  & $0.15$ & $\mathbf{0.54}$ & $0.31$  & \multirow{2}{*}{1.01} & \multirow{2}{*}{0.79} \\

 & 
 & Non-Czech  & $0.74^{*}$  & $-0.31$  & $[-3.07,\ 2.51]$  & $0.25$ & $\mathbf{0.57}$ & $0.18$  &  &  \\

 & \multirow{2}{*}{Rejection}
 & Czech  & $0.49$  & $-0.70$  & $[-3.65,\ 2.16]$  & $0.14$ & $\mathbf{0.55}$ & $0.31$  & \multirow{2}{*}{0.89} & \multirow{2}{*}{0.75} \\

 & 
 & Non-Czech  & $0.36$  & $-0.33$  & $[-3.24,\ 2.52]$  & $0.17$ & $\mathbf{0.57}$ & $0.25$  &  &  \\
\hline
\multirow{4}{*}{SES}
 & \multirow{2}{*}{Friendship}
 & High SES  & $0.52^{*}$  & $0.66^{*}$  & $[0.01,\ {1.34}]$  & $0.01$ & $0.47$ & $\mathbf{0.52}$  & \multirow{2}{*}{0.36} & \multirow{2}{*}{0.27} \\

 & 
 & Low SES  & $0.51$  & $0.72^{*}$  & $[0.09,\ {1.34}]$  & $0.01$ & $0.35$ & $\mathbf{0.64}$  &  &  \\

 & \multirow{2}{*}{Rejection}
 & High SES  & $0.47$  & $-0.68$  & $[-1.69,\ 0.31]$  & $0.03$ & $\mathbf{0.62}$ & $0.35$  & \multirow{2}{*}{0.60} & \multirow{2}{*}{0.07} \\

 & 
 & Low SES  & $0.54^{*}$  & $0.05$  & $[-0.88,\ 1.02]$  & $0.11$ & $\mathbf{0.80}$ & $0.09$  &  &  \\

\hline
\end{tabular}
\caption{Bayesian Fisher non-central hypergeometric model results. $q(\alpha)$ denotes the baseline same-group preference probability at equal in-group and out-group classroom composition. $\beta$ is the posterior median composition effect, with 95\% credible interval shown in brackets. PMP$_0=1/(1+2\cdot\text{BF}_{10}(\beta))$ gives the posterior model probability of the null under equal prior model weights, whereas PMP$_\text{contact}$ and PMP$_\text{conflict}$ partition the remaining posterior mass between the two directional alternatives. Boldface marks the highest-probability direction in each row. $\sigma_\text{class}$ and $\sigma_\text{student}$ are posterior medians of the classroom-group and student-level standard deviations. Superscript significance marks q($\alpha$) when the 90\% credible interval excludes zero and marks $\beta$ according to posterior sign probability: $^*{>}0.95$, $^{**}{>}0.99$, $^{***}{>}0.999$.
}
\label{tab:bayes_fisher_asterisks}
\end{table}

\subsection{Interpretation across the three models}\label{SI_sec:comparison-interpretation}

Taken together, the comparison helps separate behavioural from modelling effects. The binomial model shows what happens when without-replacement dependence is ignored: many qualitative patterns remain, but composition effects tend to be diluted. The Fisher model restores without-replacement sampling, but with a different interpretation of the bias parameter than in Wallenius. The fact that the strongest gender-friendship result remains visible across specifications increases confidence that this pattern is not an artifact of one particular likelihood. At the same time, the Wallenius model remains our specification because it matches the sequential nomination process most closely and gives the most natural interpretation of individual preference under finite-pool depletion.

Information-criterion comparisons lead to similar conclusions (Table~\ref{tab:model_fit_comparison}). For comparability, all three models in this table were fitted with the same frequentist maximum-likelihood approach described in Section~\ref{SI_sec:multilevel}, rather than with the Bayesian estimation strategy used for the posterior summaries reported above. For both gender outcomes, Wallenius provides the best fit by both AIC and BIC, whereas for ethnicity and SES the Fisher specification is marginally preferred to Wallenius and clearly preferred to the binomial approximation. Across all six combinations, the binomial model ranks last by both criteria, consistent with the substantive interpretation that ignoring without-replacement dependence loses information about the nomination process. 

\begin{table}[htbp]
\centering
\small
\begin{tabular}{lllrrr}
\hline
Attribute & Link type & Model & LogLik & AIC & BIC \\
\hline
\multirow{6}{*}{Gender}
& \multirow{3}{*}{Friendship} & Wallenius & $-4581.1$ & $9174.3$ & $9211.5$ \\

& & Fisher & $-4624.3$ & $9260.6$ & $9297.8$ \\

& & Binomial & $-5433.7$ & $10879.4$ & $10916.6$ \\

& \multirow{3}{*}{Rejection} & Wallenius & $-3686.7$ & $7385.5$ & $7421.6$ \\

& & Fisher & $-3690.9$ & $7393.8$ & $7430.0$ \\

& & Binomial & $-3721.8$ & $7455.6$ & $7491.7$ \\
\hline
\multirow{6}{*}{Ethnicity}
& \multirow{3}{*}{Friendship} & Wallenius & $-893.2$ & $1798.4$ & $1826.7$ \\

& & Fisher & $-883.9$ & $1779.7$ & $1808.0$ \\

& & Binomial & $-951.6$ & $1915.3$ & $1943.6$ \\

& \multirow{3}{*}{Rejection} & Wallenius & $-681.1$ & $1374.3$ & $1402.0$ \\

& & Fisher & $-680.3$ & $1372.6$ & $1400.4$ \\

& & Binomial & $-689.5$ & $1391.1$ & $1418.8$ \\
\hline
\multirow{6}{*}{SES}
& \multirow{3}{*}{Friendship} & Wallenius & $-3340.3$ & $6692.6$ & $6727.7$ \\

& & Fisher & $-3340.3$ & $6692.5$ & $6727.6$ \\

& & Binomial & $-3614.9$ & $7241.8$ & $7276.9$ \\

& \multirow{3}{*}{Rejection} & Wallenius & $-2255.8$ & $4523.5$ & $4557.4$ \\

& & Fisher & $-2255.6$ & $4523.2$ & $4557.1$ \\

& & Binomial & $-2340.5$ & $4693.0$ & $4726.9$ \\
\hline
\end{tabular}
\caption{Model-fit comparison across the three likelihood specifications for each attribute--tie-type combination. Lower AIC and BIC indicate better relative fit within a given combination. All models converged successfully. Values are rounded to one decimal place.}
\label{tab:model_fit_comparison}
\end{table}

\subsection{Choice of priors}\label{SI_sec:priors}

\begin{figure}[h]
    \centering
    \includegraphics[width=0.8\linewidth,trim=0cm 0cm 0cm 0.18cm,clip]{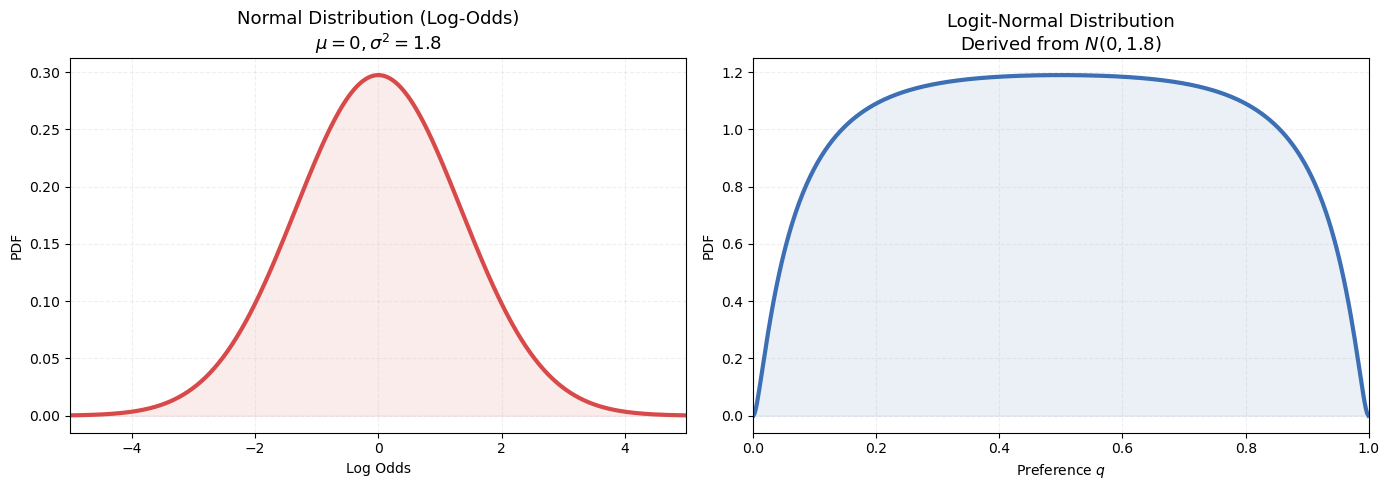}
    \caption{Prior distribution for the baseline preference parameter $\alpha$. The left panel shows the prior on the logit scale, centered at indifference, and the right panel shows the same prior transformed to the preference-probability scale. The logit-normal prior is centered at $0.5$ and places most mass away from the boundaries, favouring moderate baseline same-group preference while still allowing substantial uncertainty.}
    \label{fig:prior_alpha}
\end{figure}

For both the binomial and Fisher models, we used weakly informative priors on the baseline preference parameter $\alpha$ and the composition coefficient $\beta$:
\[
\alpha \sim \text{logit-normal}(0,1.8)
\qquad \text{and} \qquad
\beta \sim \mathcal{N}(0,16).
\]
The prior on $\alpha$ is centered at neutrality on the logit scale and, after transformation to the probability scale, places most mass on interior values rather than near the boundaries at $0$ and $1$. This reflects the idea that extreme baseline same-group preference or aversion is possible but a priori less likely than moderate values (Figure~\ref{fig:prior_alpha}).

\begin{figure}[h]
    \centering
    \includegraphics[width=0.8\linewidth,trim=0cm 0cm 0cm 0.18cm,clip]{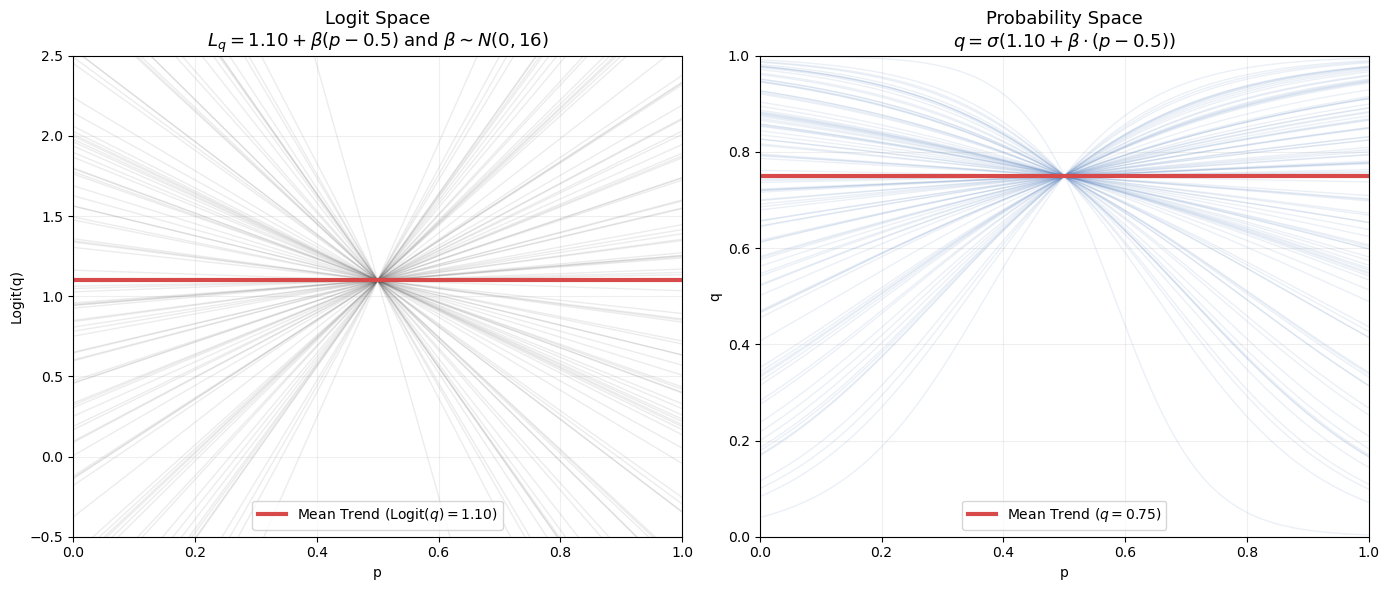}
    \caption{Implied prior variation for the composition coefficient $\beta$ with $\alpha=1.1$ fixed. The left panel shows the prior on the logit scale, and the right panel shows the corresponding curves on the preference-probability scale. The normal prior permits both positive and negative composition effects and allows a wide range of slope magnitudes, corresponding to both contact-consistent and conflict-consistent patterns.}
    \label{fig:prior_beta}
\end{figure}

The prior on $\beta$ is intentionally broad. Because $\beta$ governs how preference changes with classroom composition, a diffuse normal prior allows both contact-consistent and conflict-consistent effects, including fairly steep slopes, while still mildly regularising implausibly extreme values. Figure~\ref{fig:prior_beta} illustrates the range of composition-response curves implied by this prior.

\end{document}